\documentclass[fleqn,10pt]{wlscirep}
\usepackage{amssymb}
\graphicspath{ {./images/} }
\usepackage{stmaryrd}
\usepackage{textcomp}
\usepackage{amsfonts}
\usepackage{amsmath}
\usepackage{amssymb}
\usepackage{amstext}
\usepackage{accents}
\usepackage{epstopdf}
\usepackage{bm}
\usepackage{xfrac}
\usepackage{units}
\usepackage{textcomp}
\usepackage{multirow}
\usepackage{xr}
\usepackage{algpseudocode}
\usepackage{algorithm}
\usepackage{color}
\usepackage{xcolor}
\usepackage{subfiles}
\RequirePackage{mathptmx}
\usepackage{pgfplots}
\usepackage{pgfplotstable}
\definecolor{markercolor}{RGB}{124.9, 255, 160.65}
\pgfplotsset{
compat=1.3,
width=10cm,
tick label style={font=\small},
label style={font=\small},
legend style={font=\small}
}

\usetikzlibrary{calc}
\usetikzlibrary{intersections}

\usepackage{epsfig}
\usepackage {float}
\usepackage{graphicx}
\usepackage{geometry}
\usepackage{hyperref}
\usepackage{color,soul}
\usepackage{xcolor}
\usepackage{subfig}

\sethlcolor{yellow}
\title{Learning two-phase microstructure evolution using neural operators and autoencoder architectures}
\author[1]{Vivek Oommen}
\author[2]{Khemraj Shukla}
\author[2]{Somdatta Goswami}
\author[3,*]{R\'emi Dingreville}
\author[1,2,*]{George Em Karniadakis}
\affil[1]{School of Engineering, Brown University}
\affil[2]{Division of Applied Mathematics, Brown University}
\affil[3]{Center for Integrated Nanotechnologies, Sandia National Laboratories, Albuquerque, NM, 87185, USA}

\affil[*]{rdingre@sandia.gov, george\_karniadakis@brown.edu}


\begin{abstract}
Phase-field modeling is an effective but computationally expensive method for capturing the mesoscale morphological and microstructure evolution in materials.
Hence, fast and generalizable surrogate models are needed to alleviate the cost of computationally taxing processes such as in optimization and design of materials.
The intrinsic discontinuous nature of the physical phenomena incurred by the presence of sharp phase boundaries makes the training of the surrogate model cumbersome.
We develop a framework that integrates a convolutional autoencoder architecture with a deep neural operator (DeepONet) to learn the dynamic evolution of a two-phase mixture and accelerate time-to-solution  in predicting the microstructure evolution.
We utilize the convolutional autoencoder to provide a compact representation of the microstructure data in a low-dimensional latent space. DeepONet, which consists of two sub-networks, one for encoding the input function at a fixed number of sensors locations (branch net) and another for encoding the locations for the output functions (trunk net), learns the mesoscale dynamics of the microstructure evolution from the autoencoder latent space. The decoder part of the convolutional autoencoder then reconstructs the time-evolved microstructure from the DeepONet predictions. The  trained DeepOnet architecture can then be used to replace the high-fidelity phase-field numerical solver in interpolation tasks or to accelerate the numerical solver in extrapolation tasks.

\end{abstract}

\begin{document}
\renewcommand{\refname}{REFERENCES}
\flushbottom
\maketitle
\thispagestyle{empty}
\section*{INTRODUCTION}
\label{sec:intro}
 The phase-field method has emerged as a powerful, heuristic tool for modeling and predicting mesoscale microstructural evolution in a wide variety of material processes \cite{chen2002phase,moelans2008introduction,yang2021deep,pokharel2021physics,chakraborty2019surrogate}.
This method models interfacial dynamics without the overhead of resorting to advanced interfacial tracking algorithms such as level-set~\cite{olsson2005conservative} or adaptive meshing~\cite{yue2006phase}.
Scalar, auxiliary, continuous field variables (so called phase-field variables) are used to represent the evolutionary state of the microstructure dynamics such as in crack growth and propagation \cite{bharali2021robust,goswami2020adaptive}, thin-film deposition \cite{stewart2020microstructure, powers2021compositionally}, and dislocation dynamics \cite{beyerlein2016understanding} to name a few.
The Cahn-Hilliard, nonlinear diffusion equation~\cite{elliott1987numerical, chen2002phase, cahn1958free}, is one of the most commonly used governing equations in phase-field models.
It describes the process of phase separation, by which a two-phase mixture spontaneously separates and form domains pure in each component.
The Cahn–Hilliard equation finds applications in diverse fields ranging from complex fluids to soft matter and serves as the starting point of many phase-filed models for microstructure evolution.

Traditional numerical approaches to solve the fourth-order parabolic Cahn–Hilliard equation include 
finite differences \cite{sun1995second},
spectral approximation \cite{liu2003phase},
finite element analysis with mixed methods \cite{barrett1999finite}, and
isogeometric analysis \cite{gomez2008isogeometric,chan2019strong}.
The coupled stiff equation simultaneously captures a quick phase separation and a very slow coalescence.
Evidently, the two sub-processes operate on significantly different spatial and temporal scales, making it challenging to solve efficiently and accurately within realistic time constraints and reasonable computational capabilities \cite{alikakos1994convergence}.
Improvements in computational complexity have been enabled by the growing interest in data-driven models using machine learning (ML) methods.
However, striking a balance between computational efficiency and accuracy has often been a challenge while employing these methods.
Indeed, for complex and multi-variate phase-field models, the efficient Green's function \cite{brough2017extraction} does not ensure an accurate solution, while Bayesian optimization \cite{pfeifer2018optimization,teichert2019machine} techniques solve such coupled models but to the detriment of a higher computational cost.

Modern ML models have paved the way for the development of fast emulators for solving parametric partial differential equations (PDEs) \cite{lin2021seamless, zhang2020machine, goswami2021physics,li2020fourier, goswami2020transfer, teichert2019machine, kunselman2020semi, haghighat2021physics, zhang2020high, goswami2020adaptive_dem, attari2022machine, samaniego2020energy, raj, perera2021graph}.
A new generation of strategies for accelerating the simulation of PDEs is emerging.
A promising approach for accelerating the predictions of phase field-based microstructure evolution problems consists of using recurrent neural networks (RNNs) to learn the time-dependent, microstructure evolution in latent space\cite{de2021accelerating, hu2022RNN}.
Within this framework, statistical functions combined with linear and nonlinear embedding techniques are used to represent the microstructure evolution in latent space.
Such RNN-based surrogate models demonstrated success in generating rapid predictions of the  time evolution of the microstructural auto-correlation function.
The microstructure reconstructed from these statistical functions, using for instance a phase recovery algorithm \cite{fullwood2008microstructure}, was then used as an input for a high-fidelity solver that marches ahead in time.
The developed approach reported $5\%$ loss in accuracy against the high-fidelity phase-field solvers.
However, this class of models also come with challenges.
First, the training and inference using RNNs as a surrogate model can be relatively slow due to the temporal dependence of the current predicted field on fields predicted at previous time steps, prohibiting efficiency of the algorithm for large datasets.
Second, the RNN-based architecture learns the underlying evolution dynamics in terms of statistical functions (non-primitive variables) of the microstructure.
Reconstructing a microstructure from these statistical functions is a non-trivial and ill-posed problem \cite{herman2020data}.
This reconstruction step can incur additional errors especially for interfacial dynamics problems where resolving intricate spatial length scales such as in dendrite growth phase-field problems is key.

In this work, we propose an alternative approach to circumvent the aforementioned challenges.
We formulate the microstructure evolution problem as being equivalent to learning a mapping function $\mathcal{G}: \bm{u} \rightarrow \bm{\phi}$ such that,
\begin{equation}
    \mathcal{G}\left(  \bm{u}(x,y,t) \right) =  \bm{\phi}(x,y,t),
\end{equation}
where $\bm{u}$ is the history of  the microstructure evolution and  $\bm{\phi}(x,y,t)$ is the state of the microstructure at time $t$.
We develop a framework that integrates a convolutional autoencoder architecture with a \underline{Deep} \underline{O}perator \underline{Net}work \cite{lu2021learning} (DeepONet) to learn this mapping.
Figure \ref{fig: complete_algo} illustrates the complete end-to-end workflow of the proposed algorithm.
We utilize a convolutional autoencoder to provide a compact representation of the microstructure data in a low-dimensional, latent space.
This convolutional autoencoder approach is then combined with the DeepONet architecture to learn the dynamics of two-phase microstructures in the autoencoder latent space.
The DeepONet architecture has demonstrated its ability to model the governing differential equations (ordinary differential equations (ODEs) and PDEs) of such problems by learning the underlying operator, a mapping from functions to functions, from the available datasets for broad range of problems \cite{goswami2020transfer, lin2021operator}.
We show that such architecture is more robust than the RNN-based architecture in terms of
training,
computational efficiency, and
sensitivity to noise.
The decoder part of the convolutional autoencoder can efficiently reconstruct the time-evolved microstructure from the the DeepONet predictions bypassing the challenges associated with reconstruction-induced errors when using statistical functions to represent the microstructure for instance.
 Overall, the trained autoencoder--DeepONet framework can then be used to replace the high-fidelity phase-field numerical solver in interpolation tasks for parameters inside the distribution of inputs used during training or to accelerate the numerical solver in extrapolation tasks for parameters outside this distribution.


\section*{RESULTS}
\label{sec:results}

\subsection*{Training and optimization of neural operators and autoencoder architectures}
\label{subsec: Expt 1}

We first investigated the impact of the size of the latent dimension of the autoencoder, $l_{\rm d}$, on the model performance.
To this end, we trained five autoencoder models with $l_{\rm d} =$ 9, 25, 64, 100, and 196 respectively.
 Details of the hyper-parameters used in these five convolutional autoencoders are provided in Table~\ref{tab:conv_encoder_detail}.
For any given time step during the evolution of the microstructure, the encoder reduced a $128 \times 128$ microstructure ${\phi}(x,y,t)$ to a latent vector of size $l_{\rm d}$.
The decoder mapped the microstructural latent space representation back to a $128 \times 128$ microstructure $\hat{\phi}(x,y,t)$ (see Methods for more details).
Each autoencoder training took approximately 33 hours on one NVIDIA GeForce RTX 3090 GPU.
Next, we trained the DeepONet model for 120,000 epochs on the latent space learnt by the convolutional encoder for each of the five trained autoencoder models.
The last layer of the branch and trunk networks for all the models uses a linear activation function.
The output of the DeepONet model  was then sent to the trained convolutional decoder, which  performed a mapping from the latent space back to the original microstructure space, $\hat{\phi}(x,y,t)$.

We evaluated  the effect of the size of the latent dimension of each of the models on the basis of the relative $L^2$ norm computed across the training and testing dataset for all the time steps, including the forecasting time frames, $t= \{t_{90}, \ldots, t_{99}\}$ not seen by the surrogate model (Note that one time frame is equal to 500,000 time steps, $t=500,000\Delta t$, see Methods for additional details).
All the details of this survey analysis, including the DeepONet architecture, the $L^2$ norm of relative error on train and test datasets, and the computational time taken for training the DeepONet model are reported in Table \ref{table: latent_dim}.
From this survey, we observe that the model predictions improve when we increase the size of the latent dimension.
In general, DeepONet models with $\tanh$ and $\sin$ activation functions performed better compared to models with a ReLU activation for this particular class of problems.
As such, our best model consists of convolutional autoencoder with $l_{\rm d} = 196$ and a DeepONet model with architecture 1 and $\sin$ activation function (shown in \autoref{table: latent_dim}).
Although the training dataset consists of 1,600 different microstructure-evolution trajectories, each represented over 80 snapshots from $t=\{t_{10}, t_{11}, \ldots, t_{89}\}$, the DeepONet training is faster compared to popular RNN architectures such as the Long Short-Term Memory (LSTM) or Gated Recurrent Unit (GRU) networks\cite{lin2021operator, hu2022RNN, osorio2022forecasting}.
Since DeepONet does not have recurrent connections, there are no temporal dependencies during the training or at the inference stage.
Instead, the network relies on the convolution operations that encode information about the history through the branch network. 
In addition, due to the lack of temporal dependencies, the fully connected layers in the trunk network and convolutional layers in the branch network of the DeepONet architecture can be easily parallelized unlike LSTMs.
This makes training and inference of DeepONet significantly faster than the RNN architectures.

We carried out additional simulations to analyze the sensitivity of the proposed approach to the number of samples used for training.
We considered training datasets with 25\%, 50\%, 75\% and all the 1,600 training samples.
We adopted the same methodology proposed in Methods and trained separate autoencoder--DeepONet models on each of these datasets.
Details can be found in the Supplementary Note 2.
The model performance was evaluated on the basis of forecasting errors on test data, shown in \autoref{fig: train_samples}.
As expected, we observe better accuracy in the model predictions when increasing the number of training samples.
The model trained with 1,200 data samples shows similar accuracy to the best model trained with 1,600 data samples, indicating convergence of the training procedure.

 Finally, we also evaluated the effect of using different loss functions on training the autoencoder models.
Specifically, we trained various autoencoders by minimizing
$L^1$ loss,
relative $L^1$ loss,
$L^2$ loss,
relative $L^2$ loss, and
mixed loss ($L^2$ loss for the initial 5000 epochs and $L^1$ loss for the remaining epochs).
 The choice of a loss function determines the  landscape in a hyperspace for the optimizer to traverse in pursuit of global minima/best local minima and avoiding the saddle points.
For this task, we used a DeepONet model with architecture 1 (Table \ref{table: latent_dim}) on each of the learned latent microstructure data and re-transformed the DeepONet predictions using a pre-trained decoder to retrieve the microstructure.
We analyzed the model performance by computing the forecasting error, $\mathcal{D}_{\rm test}(t)$, on unseen test data,  as shown in Supplementary Note 3.
We observed that models for which the autoencoder is trained on $L^2$ loss performed better than the one which used the $L^1$ loss.
 When there are no outliers as solutions, $L_2$ loss is expected to perform better than $L_1$.
In the presence of outliers, $L^2$ squares them as compared to linear contribution in the $L_1$ norm.
Similarly, mean values of relative $L^1$ and $L^2$ are a better choice for autoencoder loss, $\mathcal{L}_{\rm ae}$, than the mean of $L^1$ and $L^2$, respectively.
 The relative loss values are always of $\mathcal{O}(1)$ and help in achieving convergence faster as the learning rate is of $\mathcal{O}(10^{-3})$.
We have observed such improvement in convergence in other problems, \textit{e.g.} electro-convection, where we had sharp interfaces and multiscale dynamics \cite{cai2021deepm}.
 Overall, all the models performed consistently well.
As such, our autoencoder architecture of choice is an autoencoder with $l_{\rm d} =$ 196 using a relative $L^2$ loss function.
Taken together, these results demonstrate not only the ability of our framework to accurately provide a compact representation of the microstructure data in a low-dimensional latent space,
but they also illustrate the robustness of the training of this framework.

\subsection*{Performance accuracy and forecasting ability}

\begin{figure}[H]
    \centerline{\includegraphics[width=\textwidth]{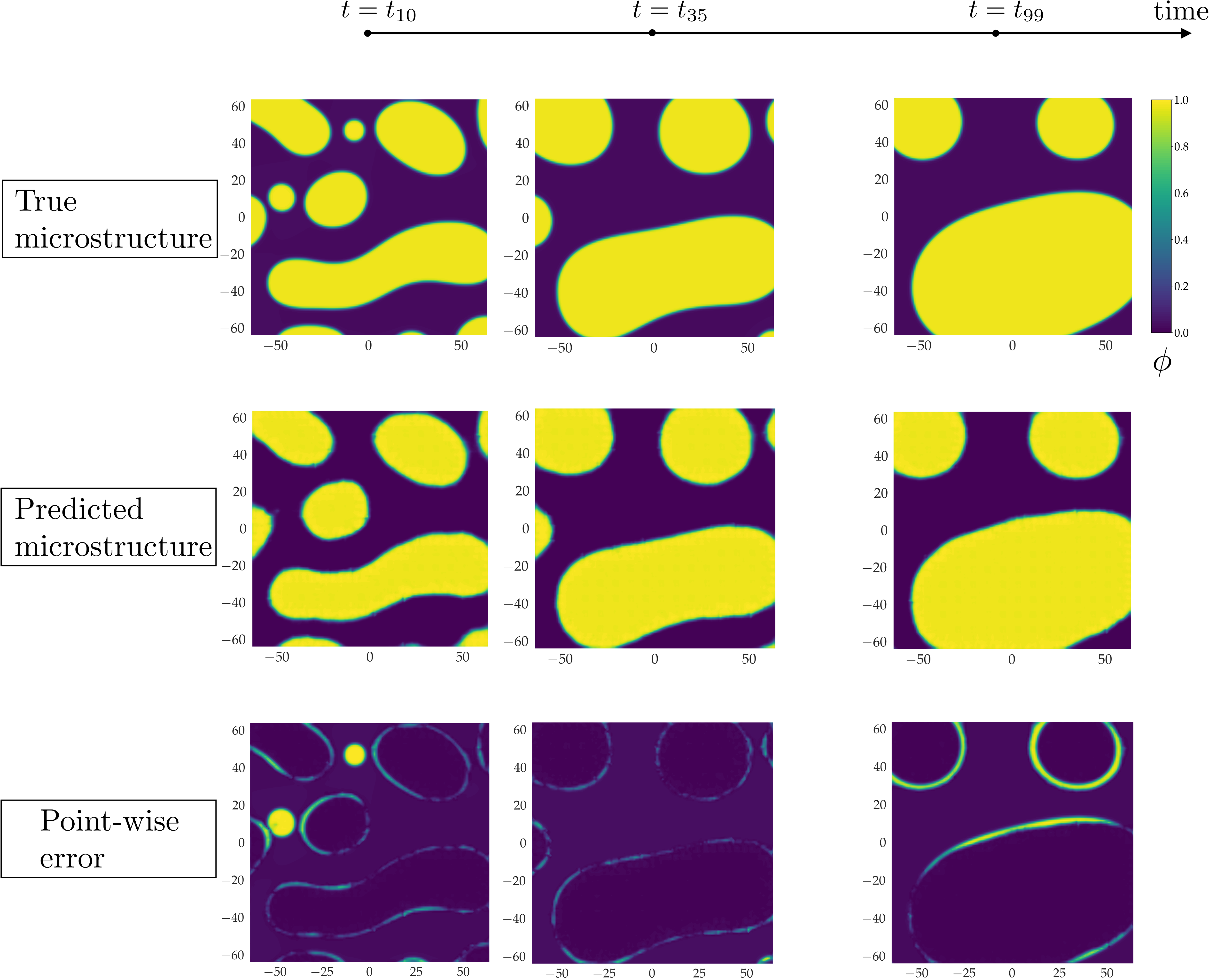}}
    \caption{{\bf Predictions of microstructure evolutions.} The true (top row), predicted (middle row), and point-wise error (bottom row) for a microstructure realization evolving in time.
    The snapshots at time frames $t = t_{10},\;t_{30},\;t_{99}$ are shown here.
    The network used for this simulation has $l_{\rm d} = 196$, and uses the following network architecture:
    Branch network -- $2\times{[}\text{conv}(128,(3,3)){]} + {[}3920{]}$;
    Trunk network -- $2\times{[}100{]} + {[}3920{]}$.}
    \label{fig: predictions}
\end{figure}

A comparison of the predictions from our accelerated framework with that from high-fidelity phase-field simulations for a representative case of microstructure evolution at three different time steps is shown in Fig. \ref{fig: predictions}.
During the initial time steps, the microstructure is rich with multiple features and evolves rapidly with respect to time.
Our autoencoder--DeepONet as a surrogate model is able to successfully predict the larger features and the overall morphology of the microstructure.
The point-wise error snapshots suggest that the model 
fails to identify the relatively smaller features in the microstructure and
contains significant errors along the sharp boundaries.
In other words, the spatial gradient of the phase concentration is not as sharp as that of the true microstructure  obtained from high-fidelity, phase-field simulations.  

\begin{figure}[H]
    \centerline{\includegraphics[width=0.75\textwidth]{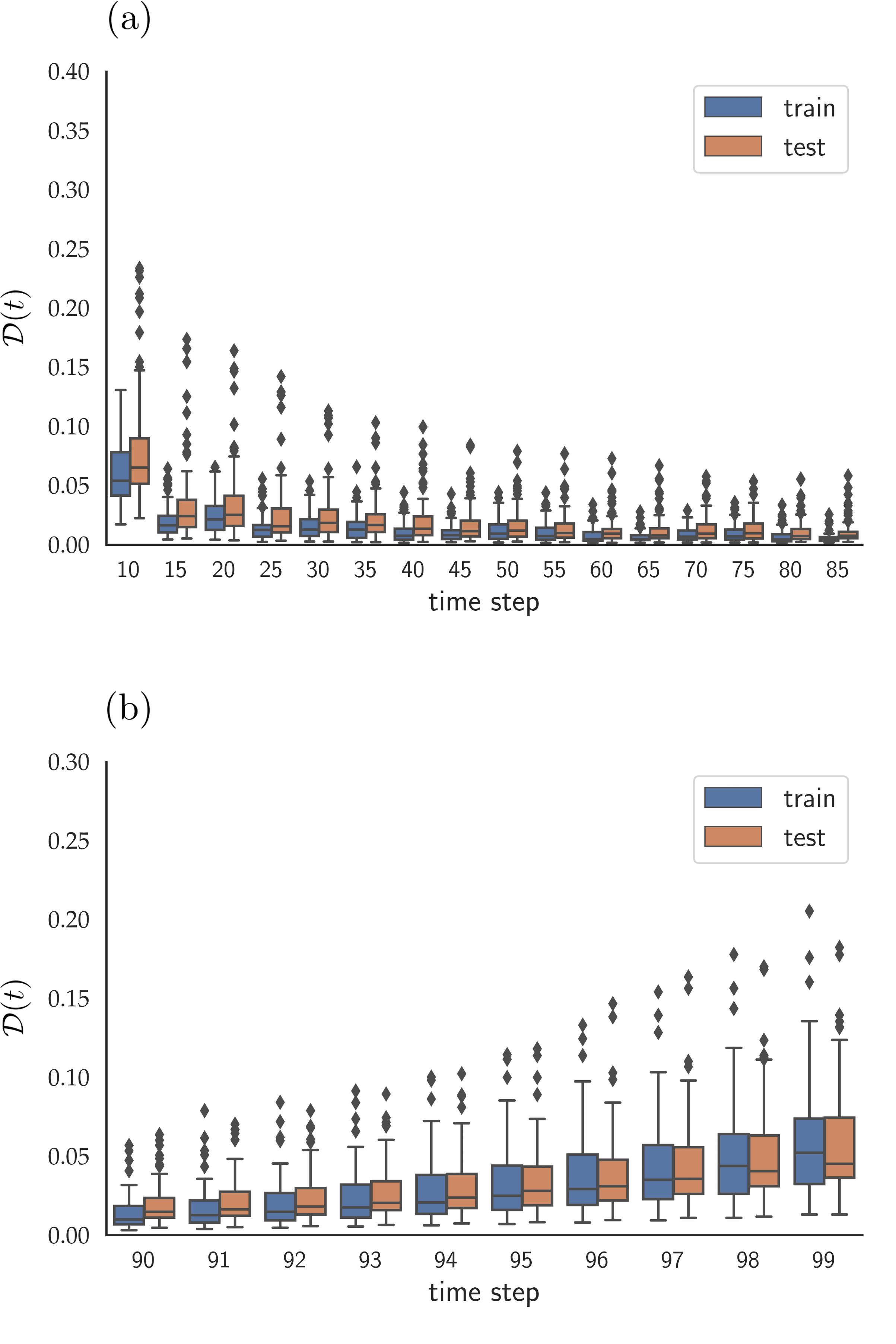}}
    \caption{{\bf $L^2$ norm of the relative error between true and predicted microstructures at each time step.}
    (a) Box plot with respect to $\mathcal{D}(t)$ computed over the training time steps over the train and test datasets.
    (b) Same error metric, but in future time steps never seen during the training phase.}
    \label{fig: error_boxlplot}
\end{figure}

From Fig. \ref{fig: predictions}, we qualitatively get the intuition that the predicted microstructures contain errors at the earlier time steps because of the missing, small-size features and at the later time steps due to smoother boundaries predicted by the model.
To confirm and quantify this notion, we computed the $L^2$ norm of the relative error at each time step, $\mathcal{D}(t)$, defined as:
\begin{equation}
    \label{eqn: l2 norm(t)}
    \mathcal{D}(t) = \frac{ \sum_x \sum_y \left( \phi(x,y, t) - \hat{\phi}(x,y, t; \bm{\theta}) \right)^2}{ \sum_x \sum_y \phi(x,y,t)^2},~t\in\left[t_{10},\;t_{11},\;\ldots,\;t_{99}\right].
\end{equation}
To analyze the accuracy of the prediction at each time step, we calculated $\mathcal{D}(t)$ across the samples in the training and testing datasets, and created a boxplot as shown in Fig. \ref{fig: error_boxlplot}.
The error is high for the initial time steps, where features span multiple length scales and evolve rapidly with time.
However, the predictions improve over time when the evolution process slows down and the microstructure features coarsen.
The time steps shown in Fig. \ref{fig: error_boxlplot}(a) were used during the training of the model.

Next, we evaluated the capability of the model to forecast time frames $t=\{t_{90}, t_{91}, \ldots, t_{99}\}$.
From Fig. \ref{fig: error_boxlplot}(b), the error is seen to increase gradually when the model extrapolates at unseen time instances.
A closer look at the forecasting predictions offers further insights on the DeepONet predictive performance.
We computed the mean of $\mathcal{D}(t)$ across the training and testing datasets for all the models given in Table \ref{table: latent_dim}.
We also plotted these values in \autoref{fig: forecasting_error} with additional details in the Supplementary Note 4.
We observe that the mean relative $L^2$ error reduces when increasing the latent dimension of the autoencoder model.
In other words, the model with a larger latent space is able to predict the evolution of the microstructure in forecasting mode.
This is intuitive because a larger dimension of the latent space implies that there are more basis functions to express the encoded information about the microstructure and its evolution, and therefore the network has an improved representation capability.
However, this trend seems to saturate beyond $l_{\rm d} = 100$.
For the model with $l_{\rm d} = 100$ or $l_{\rm d}=196$, the forecasting error is always less than $6\%$.
The logarithm of the relative $L^2$ error linearly increases for $l_{\rm d}=64,\;100$ and $196$ for forecasting time step, whereas for $l_{\rm d}=9$ and $l_{\rm d}=25$, the error is high and remains constant.

\subsection*{Robustness of the surrogate DeepONet framework: Sensitivity to noise}

We evaluated the accuracy and robustness of the predictions from our surrogate model by systematically increasing the noise levels in the model input.
For this analysis, we considered the best model with $l_{\rm d}=196$ and DeepONet architecture 1 (see Table \ref{table: latent_dim}) with $\sin$ activation functions.
We added Gaussian white noise to our microstructure data with zero mean and standard deviations, $\sigma = 0.5\%, 1\%, 2\%, 3\%, 4\%, 5\%, 10\%$.
To evaluate the model performance, we used the relative $L^2$ norm, $\mathcal{D}$, as defined in Eq.~\eqref{eqn: l2 norm}.
 The forecasting error, $\mathcal{D}$, is calculated across the samples present in the test dataset.
Details can be found in the Supplementary Note 5.

From \autoref{table: noise} and \autoref{fig: sensitivity_to_noise}, the relative $L^2$ norm does not increase noticeably when noise is added to the model input.
 In fact, the surrogate is almost invariant to noise up to 10\% Gaussian white noise, as  presented in \autoref{fig: sensitivity_to_noise}.
Previous studies \cite{vincent2008extracting, vincent2010stacked, gondara2016medical} illustrated the capability of autoencoders to denoise noisy images.
The transformation to a low-dimensional latent space forces the autoencoder to retain the dominant features alone while discarded unnecessary noise.
The convolutional autoencoder used in our approach does exactly that by denoising the noisy microstructure input.
 The encoder filters out noise and only retains the dominant energy modes of the microstructure data.
The output of the convolutional encoder is almost in its pure form, free from noise and therefore it enables the DeepONet to make stable predictions.
The decoder accurately reconstructs the microstructure from the predictions made by DeepONet in the latent space.
This performance illustrates the robustness and efficacy of the present framework as compared to other machine-learned frameworks that use statistical functions to encode the microstructure representation~\cite{herman2020data}.
Indeed, it has been illustrated by Herman and coworkers~\cite{herman2020data} and others~\cite{bostanabad2018computational} that, while statistical functions such as the microstructure auto-correlation functions are sufficient to capture the salient features of the microstructure in latent space, such representation does not uniquely map back to the true microstructure as it is an ill-posed, inverse problem.
Here, our results essentially bypass such challenge by taking advantage of the fact that autoencoders are robust to corruption in the representations they learn. 
 The denoising nature of a trained autoencoder enables the encoder to learn a stable and consistent mapping to the latent space.
This makes training of the DeepONet much more stable and results into accurate predictions of the microstructure at any desired time step.

\subsection*{Effect of time resolution}

The high-fidelity phase-field forward numerical solver (MEMPHIS) discretizes the time with a time step $\Delta t = 1\times10^{-4}$ (see Methods for additional details).
Stability of this numerical integration scheme can be achieved by strictly following the Courant–Friedrichs–Lewy (CFL) condition to solve the Cahn-Hilliard equation for 50M time steps.
The solver saves snapshots of the solution at every 500,000$^{\text{th}}$ time step resulting in 100 microstructure time frames for each realization.
We initially utilized 80 equally spaced time frames between the $10^{\rm th}$ and $90^{\rm th}$ time frames for training the surrogate model.
Therefore, for the surrogate DeepONet model, each time step was $ 500000 \times \Delta t = 50$.

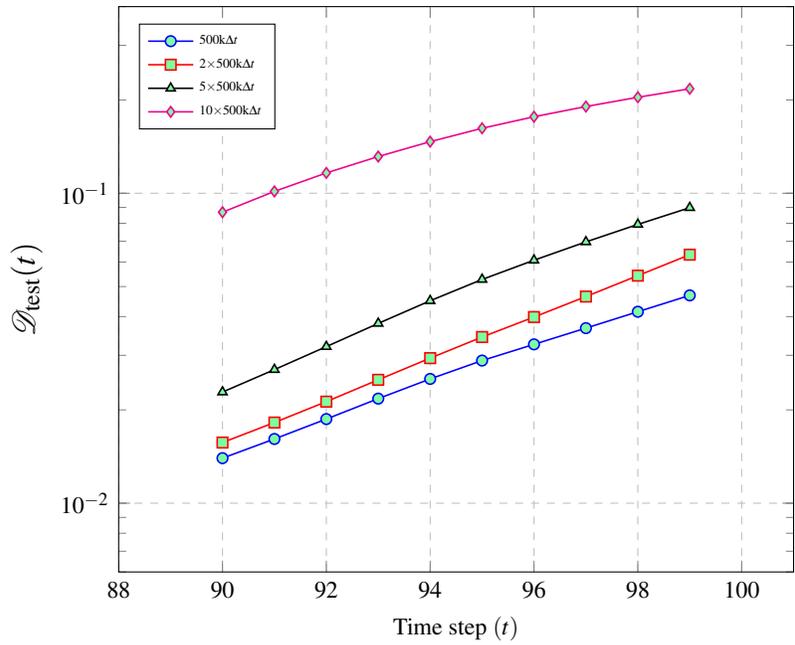
\begin{figure}[H]
	\centering
		\begin{tikzpicture}
		\begin{semilogyaxis}[
		legend cell align=left,
		width=.6\textwidth,
		xlabel={Time step $(t)$},
		ylabel={\large{$\mathcal{D}_{\rm test}(t)$}}, 
		xmin=88, xmax=101,
		ymin=6e-03, ymax=4e-1,
		legend pos=north west, legend cell align=left, legend style={font=\tiny},
		xmajorgrids=true,
		ymajorgrids=true,
		grid style=dashed,
		] 	
		\addplot[color=blue,mark=*,semithick, mark options={fill=markercolor}]
		coordinates{(90,0.013973691) (91,0.016115785) (92,0.018685032) (93,0.021745443) (94,0.025173184) (95,0.028853958) (96,0.03254943) (97,0.036718503) (98,0.041472007) (99,0.046830278) };
		
		\addplot[color=red,mark=square*,semithick, mark options={fill=markercolor}]
		coordinates{(90,0.015692737) (91,0.018209841) (92,0.021263894) (93,0.025007904) (94,0.029376779) (95,0.034371283) (96,0.039871637) (97,0.046418153) (98,0.054219104) (99,0.06335973) };
		
		 \addplot[color=black,mark=triangle*,semithick, mark options={fill=markercolor}]
		coordinates{(90,0.0228323) (91,0.026975695) (92,0.032002278) (93,0.038029633) (94,0.0449938) (95,0.052707028) (96,0.06080662) (97,0.06966974) (98,0.07934471) (99,0.08982988) };
		
		\addplot[color=magenta,mark=diamond*,semithick, mark options={fill=markercolor}]
		coordinates{(90,0.08686262) (91,0.10138206) (92,0.116268456) (93,0.13135095) (94,0.14670846) (95,0.16203776) (96,0.17653169) (97,0.19048378) (98,0.20402817) (99,0.21728575) };

		\legend{500k$\Delta t$, 2$\times$500k$\Delta t$, 5$\times$500k$\Delta t$, 10$\times$500k$\Delta t$}
		\end{semilogyaxis}
		\end{tikzpicture}
	\caption{{\bf Variation of forecasting error on test data for models trained on data with different spacing in time.} Spacing in time tested are: 500k$\Delta t$, 2$\times$500k$\Delta t$, 5$\times$500k$\Delta t$, 10$\times$500k$\Delta t$. $\mathcal{D}_{\rm test}(t)$ represents the relative $L^2$ error computed across the samples in test dataset at different time steps. }
	\label{fig: time_spacing}
\end{figure}

To investigate the effect of different spacing of physical time on the surrogate DeepONet model, we performed a sensitivity study using data that is spaced differently in time to train the model.
Specifically, we trained DeepONet models on datasets with time spacing of
500k$\Delta t$,
2$\times$500k$\Delta t$,
5$\times$500k$\Delta t$ and
10$\times$500k$\Delta t$.
The DeepONet predictions were then remapped to the primitive space using the pre-trained convolutional decoder to recreate the microstructure at the required time step.
We plot the mean relative $L^2$ forecasting error corresponding to the models trained on differently spaced datasets in \autoref{fig: time_spacing}.
As expected, just like with any other time-integration scheme, we observe that the forecasting error increased for larger spacing in physical time.
However, the computational efficiency of the DeepONet predictions remained the same to predict consecutive time frames regardless of the time spacing used.

\subsection*{Training strategy for learning concurrently multi-scale features}

We showed in a previous work~\cite{de2021accelerating, hu2022RNN} that the microstructure evolution in latent space is non-linear.
Indeed, for early time steps the microstructure evolves rapidly and then later on it evolves more slowly once the phase separation dynamics has taken effect.

We noted that the DeepONet model architecture presented above was not able to resolve the small scale microstructural features as shown in \autoref{fig: predictions}.
 In the early time steps, which represent fast dynamics, small wavelength features are hard to capture due to spectral bias of neural networks;
\autoref{fig: error_boxlplot} quantifies this difficulty.

\begin{figure}[H]
    \centerline{\includegraphics[width=0.8\textwidth]{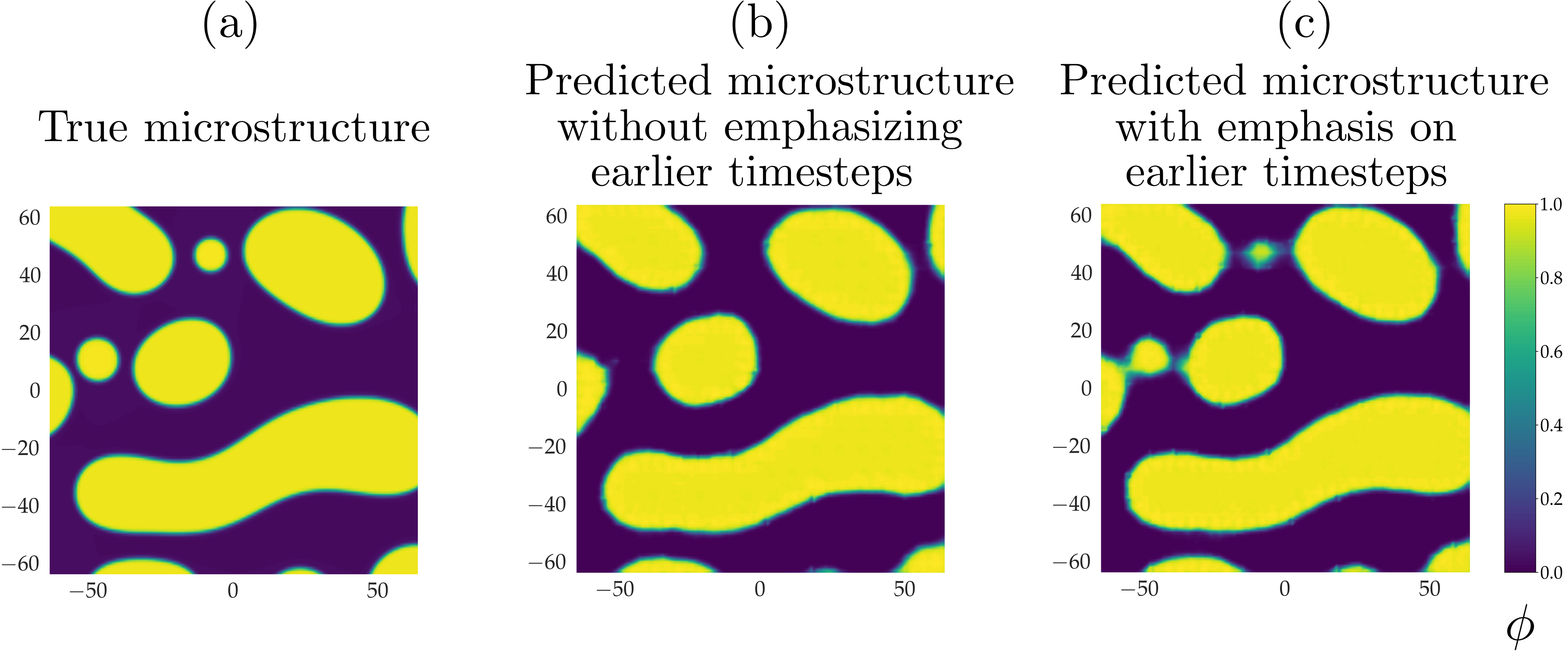}}
    \caption{{\bf Capturing small microstructural features at earlier time steps.}
    (a) True microstructure at $t=t_{10}$.
    (b) Predicted microstructure without emphasizing earlier time steps during training.
    (c) Microstructure predicted by DeepONet trained on a dataset where the earlier time steps where repeated to increase the importance given to earlier time steps.}
    \label{fig: initial_repeat}
\end{figure}

To circumvent this issue, during training of the DeepONet model, we increased the weight given to earlier time steps of each realization in the dataset.
 By placing more emphasis to early snapshots during training, we endow DeepONet with an inductive bias to learn the fast dynamic accurately.
Practically speaking, we are forcing the DeepONet model, $\mathcal{G}(\tilde{\bm{\Phi}})(t; \bm\theta_{\rm d})$, to predict earlier time steps repeatedly by creating a new training dataset with repeated $\tilde{\bm{\phi}}(t)$ for each realization.
Since the DeepONet model is trained to minimize the mean squared error between the true and predicted microstuctures, the model is driven to give greater emphasis to microstructures developing at earlier time steps.
The results from this training procedure are depicted in Fig. \ref{fig: initial_repeat}.
Indeed, from the comparison of the predicted microstructure without and with an emphasis on earlier time steps in Fig. \ref{fig: initial_repeat} (b) and (c) respectively, we observe that increasing the weight given to the earlier time steps of evolution for each realization results in a DeepONet model capable of recovering smaller, high-frequency components.
This ability to accurately resolve multiple length scales is particularly important in dynamic problems such as dendrite or grain growth problems for instance, where the simulated microstructure dynamics can be extremely sensitive to the development of multiple length scales concurrently.

\subsection*{Integration of DeepONet with a numerical high-fidelity phase-field solver}

The results above show that a pre-trained autoencoder--DeepONet model can be used as a robust and efficient surrogate of the numerical solver when inference is requested for initial microstructure and parameters within the distributions of the training datasets (interpolation task).
Our proposed framework can also be used for extrapolation tasks and be integrated to the phase-field numerical solver to accelerate the predictions for initial microstructure and parameters that are outside the aforementioned distributions (extrapolation task).
To demonstrate this point, we devised a hybrid approach that integrates the autoencoder--DeepONet framework with our high-fidelity phase-field Mesoscale Multiphysics Phase Field Simulator (MEMPHIS solver).
This hybrid model unites the efficiency and computational speed of the autoencoder--DeepONet framework with the accuracy of high-fidelity phase-field numerical solvers.

\begin{figure}[H]
    \centerline{\includegraphics[width=0.93\textwidth]{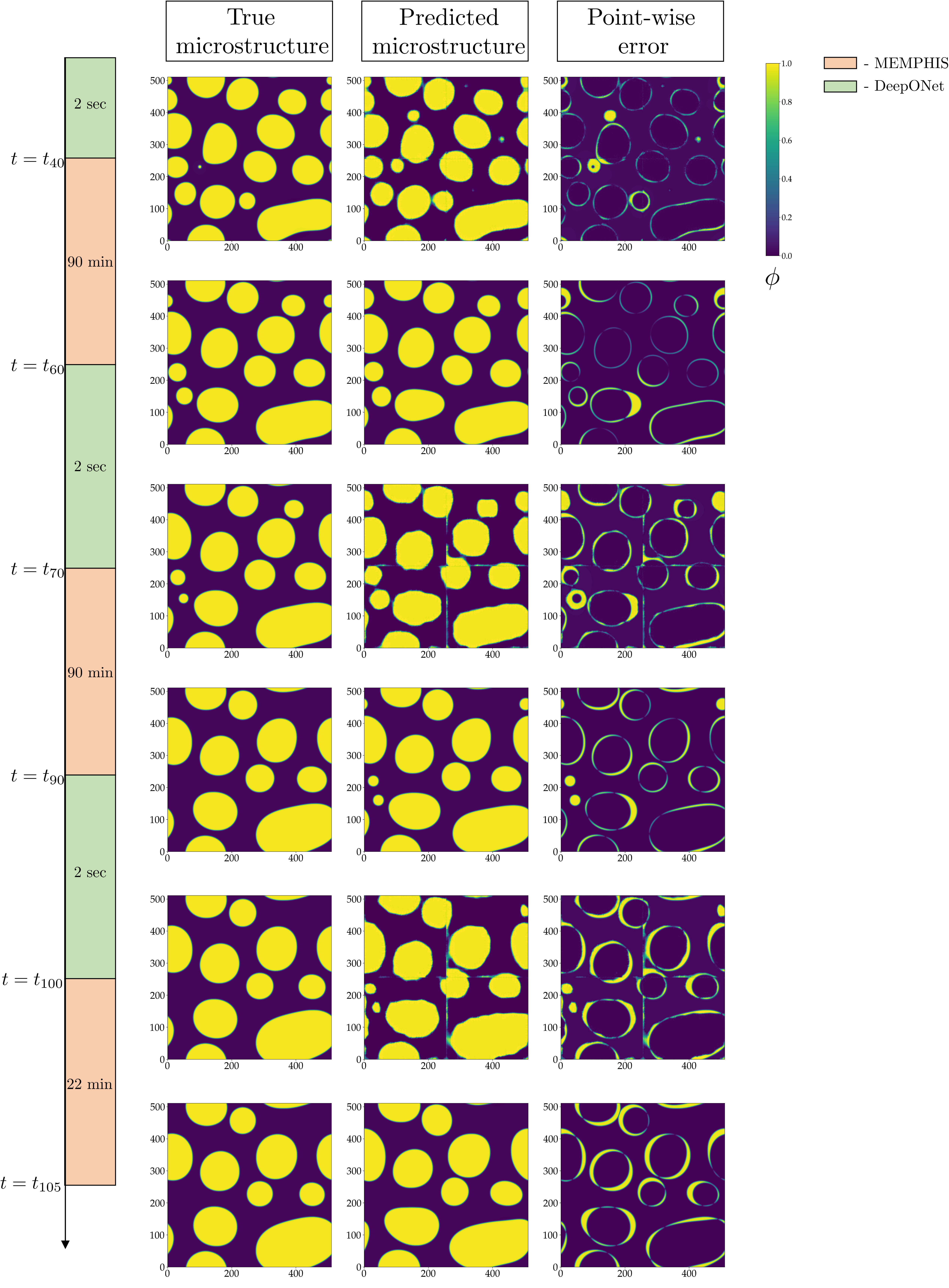}}
    \caption{{\bf Schematic of our hybrid approach for integrating DeepONet with the numerical solver MEMPHIS to accelerate phase-field predictions.}
    The computational time corresponding to the autoencoder--DeepONet model and the MEMPHIS solver for one realization in the training dataset is reported in this figure. The error is shown on the third column.}
    \label{fig: integration_results_train}
\end{figure}

The hybrid framework consists of alternating between predictions from the high-fidelity phase-field simulations and that from the autoencoder--DeepONet model.
The high-fidelity phase-field simulation step provides accuracy in the description of the dynamics, while the autoencoder--DeepONet model enables us to `leap in time'.
The algorithm is presented in Algorithm \autoref{alg: mod}.
Here we choose to split the time evolution predicted between the high-fidelity simulation and that of the autoencoder--DeepONet to be equal to one another.
Each solver within this integrated scheme sequentially predicts 10 time frames, which corresponds to $10 \times 500\text{k} = 5\text{M}$ time steps for the high-fidelity phase-field solver alone.
A schematic of the approach and results are shown in \autoref{fig: integration_results_train} for the training data.
Results for the test data are provided in the \autoref{fig: integration_results_test}.
The discontinuity along the centerlines of the predictions made by the  autoencoder--DeepONet model arises from splitting each realization into four different realizations as discussed in the `Microstructure-evolution dataset' subsection.

We see in \autoref{fig: integration_results_train} that the forecasting from 10 time frames using the high-fidelity phase-field solver MEMPHIS running on 32 CPU-cores (Intel$^{\tiny{\text{\textregistered}}}$ Xeon$^{\tiny{\text{\textregistered}}}$, e5-2670) takes approximately 90 minutes.
The subsequent 10 time frames predicted by the autoencoder--DeepONet model take only 2 seconds.
A comparison of the computational cost between the high-fidelity phase-field solver alone and the hybrid approach is reported in \autoref{table: integration}.
Here, we achieve a speed-up of 29\% .
This performance can be improved by a much greater factor with more extensive offline training with a richer data set of operating conditions, which will lead to better generalization.
For each evolution, our hybrid approach saves 135 minutes, without loss of accuracy as shown previously in the Results section.
The choice of a specific time step splitting, for which the system is evolved using the high-fidelity phase-field framework and then by the autoencoder--DeepONet model, is arbitrary and can be considered as a hyper-parameter.
For instance, one could easily consider to use very short time steps within the high-fidelity phase-field solver window to course-correct the physical predictions and much longer time steps when using DeepONet to accelerate the time evolution predictions.
This type of time splitting integration scheme would dramatically increase the speedup even further.
Additionally, although we showed the microstructure evolution only until $t=t_{105}$, such a hybrid time integration strategy can be adopted to forecast time evolution of the microstructure for time windows that can be much longer, for instance as long as the input time history used to train the DeepONet model while still keeping a good accuracy, leading to substantial savings in CPU hours.


\section*{DISCUSSION}
\label{sec:summary_and_discussions}

In this work, we investigated the effectiveness of a convolutional autoencoder--DeepONet approach for modeling the evolution dynamics of mesoscale microstructures.
  The proposed framework consists of two parts.
First, learning a non-linear mapping to a latent manifold using convolutional autoencoders, and
second, learning the dynamics in the latent space (from the first step) using DeepONet.
We trained our model on high-fidelity, phase-field data generated by solving the Cahn-Hilliard equation.
The results presented above show that the trained DeepOnet architecture can be used robustly to replace the high-fidelity phase-field numerical solver in interpolation tasks or to speed up the numerical solver for extrapolation tasks.
We showed that increasing the latent dimension used to describe the microstructure evolution and putting more emphasis on earlier time steps during the training improve the overall representation capability of the framework.
Given its performance, this framework offers several advantages as compared to other machine-learned architectures used for accelerating the prediction of the phase-field-based microstructure evolution.

\begin{figure} [H]
    \centerline{\includegraphics[width=0.95\textwidth]{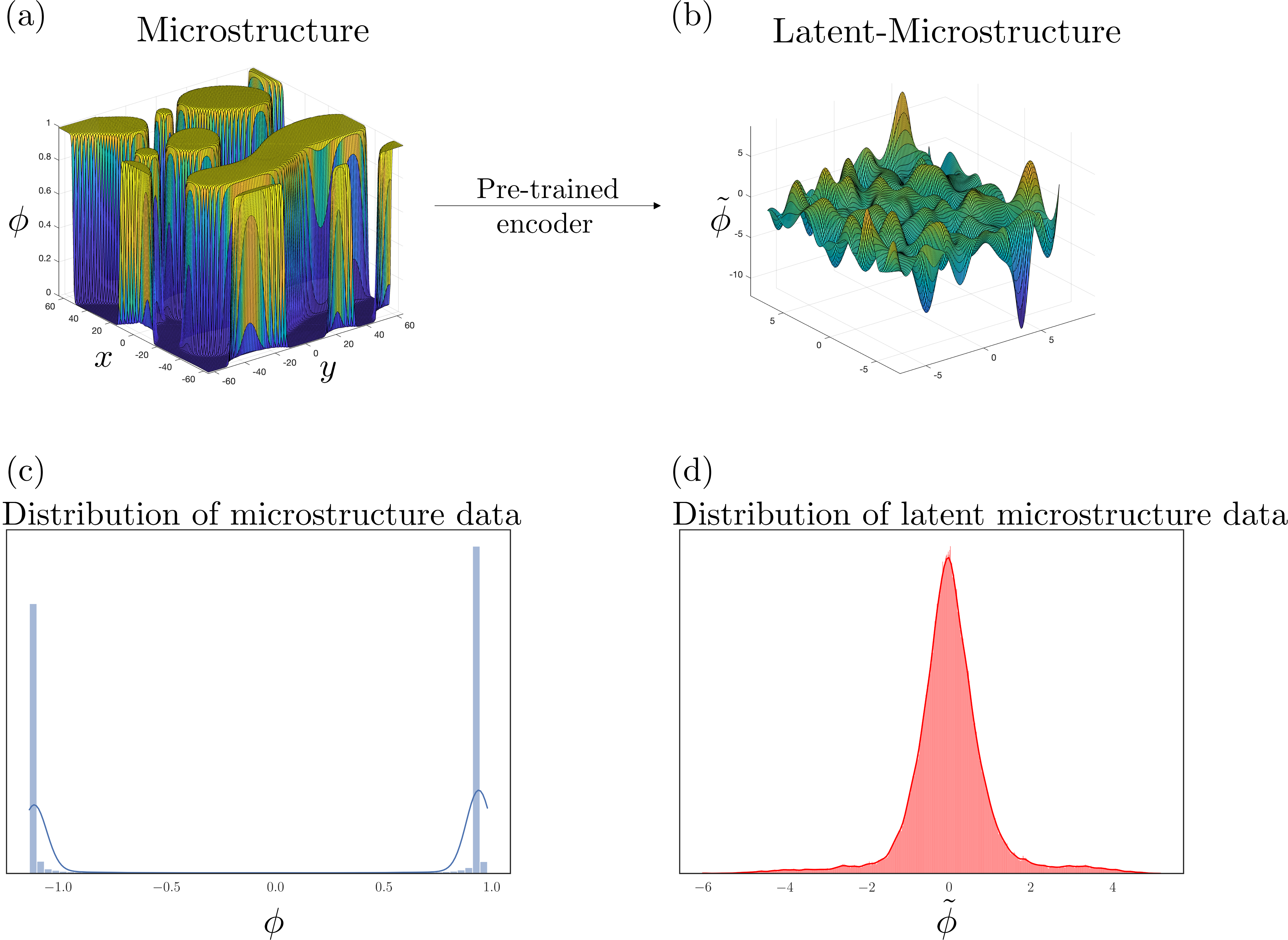}}
    \caption{ {\bf Representation of microstructure data and statistical insights.}
    (a) represents a 3D visualization of the function to be approximated by the surrogate model. The presence of several high-gradient regions at every time-step, makes it challenging for neural network models to learn the evolution dynamics of microstructures.
    Panel (b) represents a smoother latent-microstructure learned by the encoder during the autoencoder training.
    (c) The microstructure data, $\phi(x,y,t)$, is predominantly represented by 1s or 0s.
    (d) The encoder transforms $\phi$ to a latent space, $\tilde{\phi}$, where deep neural networks can learn easily. The curves in (c) and (d) represent the smoothed density estimates of the histogram}
    \label{fig: distributions}
\end{figure}

First, unlike existing methods \cite{herman2020data,de2021accelerating} that train machine-learning-based surrogate models using low-dimensional representations of microstructures based on statistical functions (\textit{e.g.} auto-correlation function), our autoencoder--DeepONet approach learns a suitable low-dimensional latent space using a convolutional autoencoder.
We showed that this approach bypasses any post-processing steps (\textit{e.g.} a phase-recovery algorithm) necessary to reconstruct the microstructure from statistical functions \cite{herman2020data}.
The advantage of training DeepONet in the autoencoder latent space is two-fold. 
On one hand, training DeepONet in a low-dimensional space is computationally efficient. 
On the other hand, the presence of high-gradient regions in the microstructure data (see Fig. \ref{fig: distributions} (a)) can make the training of the model challenging.
However, the encoder transforms microstructure data to a latent space (Fig. \ref{fig: distributions}(b)), where the gradients are not as high and more gradual.
In other words, the encoder learns a non-linear mapping of the microstructure data coming from an untrainable distribution in the primitive space, as shown in Fig. \ref{fig: distributions}(c), to a trainable distribution in the low dimensional latent space, as shown in Fig. \ref{fig: distributions}(d).
Such transformation from data with high gradient to data with gradual, smoother gradient facilitates the training of the DeepONet model.
Although we have trained our own autoencoder model in this work, we believe that fine-tuning any autoencoder pre-trained on existing image datasets with similarities to our microstructure images will be suitable for this task.
Reusing such readily available pre-trained autoencoders can further save on the computational cost of our workflow.

Second, even though the workflow presented in this work is focused on two-dimensional (2D) microstructure data, it can easily be extended to three-dimensional (3D) microstructure data.
For 3D microstructure evolution case, each realization can be represented by a sequence of 3D tensor data structures.
Hence, we could use an autoencoder with 3D convolution layers in the encoder and 3D transpose convolution layers in the decoder and learn a suitable non-linear mapping to a low dimensional latent manifold.
Following the same approach, a DeepONet model can be trained to learn the dynamics in the latent space, and be then remapped to primitive space using the already trained decoder.
As shown in a recent theoretical paper, DeepOnet can tackle the curse of dimensionality in the input space, so training it in high-dimensions is not a prohibitive issue\cite{lanthaler2022error}.

Third, the autoencoder and DeepONet are trained solely from data, making the proposed approach purely data-driven, independent of the boundary conditions.
The boundary conditions are never explicitly assumed as an input data to the framework at any stage.
They are implicitly fed through the latent representation of the microstructure history data inputted to the branch network of DeepONet.
Therefore, any information regarding a change in the boundary condition will be available in the latent microstructure history fed to the branch network, enabling the DeepONet model to predict the dynamics accordingly in the latent manifold.
In this manner, the purely data-driven nature makes the proposed autoencoder--DeepONet framework agnostic to any changes in boundary conditions within the considered history.
We also need to clarify  that periodic boundary conditions were imposed at all four boundaries of the computational domain while generating the data from the numerical solver MEMPHIS.
We note that DeepONet can be trained to map boundary conditions to an output field if so desired for a very general set of variable boundary conditions.

There are several extensions to the present framework that can be implemented in order to improve the accuracy, predictability, and acceleration performances.
These improvements are related to the training of the model and extension to a multi-fidelity implementation.
The first topic is related to improving the accuracy of the model with physics constraints in order to better capture non-linearities in the model evolution. The second topic is related to fusing different sources of data within our data set, \textit{e.g.}, `low' fidelity from simulation and `high' fidelity from physical experiments of the same nature.

Regarding a physics-informed implementation, Wang et al. \cite{wang2021learning} for instance put forward a physics-informed DeepONet, where the PDE of the underlying system is added as a soft constraint to the loss functions.
In the present study, the training framework is purely data driven and we are learning the dynamical system in the latent space defined by non-primitive coordinates except the time, which is fed as an input to the trunk net. 
However, similar to Wang et al., some physical constraints could be injected in the current framework by using for instance the fact that the mass $\phi$ is conserved at all times.

Regarding the multi-fidelity implementation, the present approach can be extended to incorporate experimental data coming from similar processes.
For instance, recently, several researchers~\cite{de2022bi, howard2022multifidelity, lu2022multifidelity} explored diverse ways of exploiting the inherent correlations between datasets coming from different sources of data with different levels of fidelity and obtaining optimal predictions.
In this context, the data obtained from phase-field models using a numerical solver can be considered as a low-fidelity dataset and the limited amounts of experimental microstructure image data from similar processes can be treated as high-fidelity dataset.
The autoencoder--DeepONet framework proposed here can be extended to generate accurate predictions from a limited number of high-fidelity experimental microstructure microscopy image data, by utilizing the high correlation with the surplus low-fidelity phase field data.
The assimilation of experimental data in the present DeepONet architecture can be concatenated with numerical data as another realization and the proposed workflow will remain unchanged.
Merging and taking advantage of both experimental and modeling efforts is a future direction of our research.

To summarize, we developed and applied a machine-learned framework based on neural operators and autoencoder architectures to efficiently and rapidly predict complex microstructural evolution problems.
Such architecture is not only computationally efficient and accurate but it is also robust to noisy data.
The demonstrated performance makes it an attractive alternative to other existing machined-learned strategies to accelerate the predictions of microstructure evolution.
It opens up a computationally viable and efficient path forward for discovering, understanding, and predicting materials processes, where evolutionary mesoscale phenomena are critical, such as in optimization and design of materials problems. 


\section*{METHODS}
\label{sec:methodology}

\subsection*{Phase-field model of the spinodal decomposition of a two-phase mixture}
\label{subsec: phase field model}

We illustrate our accelerated phase-field workflow on the simplest case of the spinodal decomposition of a two-phase mixture.
This model is highly relevant to many phase-field models.
In the spinodal decomposition of a two-phase mixture uses a single order parameter, $\phi({\bf x}, t)$ to describe the atomic fraction of solute diffusing within a matrix.
The free energy of the system is expressed by the Cahn-Hilliard equation based on the Onsager force–flux relationship such that 
\begin{equation}\label{eq:cahn_hilliard}
    \frac{\partial \phi}{\partial t} = \nabla\cdot\left(M_{\rm c}(\phi)\nabla [\omega_{\rm c}(\phi^3-\phi)+\kappa_{\rm c}\nabla^2\phi]\right),
\end{equation}
where $\omega_{\rm c}$ is the height of the energy barrier between the two phases,
$\kappa_{\rm c}$ is the gradient energy coefficient, and
$M_{\rm c}$ denotes the concentration dependent mobility, with
$M_{\rm c} = s(\phi)M_{\rm A} +(1-s(\phi))M_{\rm B}$.
The function $s$ defines a smooth interpolation to switch from phase `\textit{A}' to phase `\textit{B}'.
This interpolation function is defined as $s(\phi) = \frac{1}{4}(2-\phi)(1+\phi)^2$.
In the present model, both the mobility and the interfacial energy are taken to be isotropic and $\omega_{\rm c}$ and $\kappa_{\rm c}$ are stet to unity for simplicity.
The evolution of one phase is expressed as a symmetric double-well potential, with minima at $\phi\pm1$.

\subsection*{Microstructure-evolution dataset}
\label{subsec:data_generation}

The phase-field model described above is implemented using Sandia’s in-house multi-physics phase-field modeling code MEMPHIS~\cite{stewart2020microstructure,dingreville2020benchmark}.
In order to generate a diverse and large set of simulation results exhibiting a rich variety of microstructure features, we independently sampled the phase fraction $\phi_{A}$, such that each phase has at least a minimum concentration of $0.15$ (note that $\phi_{\rm B} = 1-\phi_{\rm A}$), and the phase mobilities $M_{\rm{A}}$ and $M_{\rm{B}}$ of species `A' and `B'.
Phase mobilities are sampled independently to vary in the range $[0.01, 100]$.
In total we generated 500 triplets $(\phi_{\rm A}, M_{\rm A}, M_{\rm B})$ using Latin Hypercube Sampling.
 In the simple case of the spinodal decomposition, only the tuple $(\phi_{\rm A}, M_{\rm A}/M_{\rm B})$ is necessary.
As demonstrated in other studies~\cite{stewart2020microstructure, herman2020data} that share similar microstructure evolution as the spinodal decomposition, it is however necessary to handle $(\phi_{\rm A}, M_{\rm A}, M_{\rm B})$ separately since the ratio $M_A/M_B$ by itself will not be sufficient anymore to characterize the dynamic of the microstructure evolution.
Herein, we frame the present work in a broader context for generality.
All the simulations were performed using a two-dimensional (2D) square domain $\Omega= [0,1] \times [0,1]$, discretized with $512\times512$ grid points, with a dimensionless spatial discretization of unity on either direction, and a temporal discretization of $\Delta t = 1\times10^{-4}$.
The simulation domain's composition field is initialized using truncated random Gaussian distribution in the range $[-1,1]$ with $\mu = \phi_{\rm A}$, and $\sigma = 0.35$.
The microstructure was allowed to evolve and grow for 50,000,000 time steps, saving the state of the microstructural domain every 500,000 time steps, hence a total of 100 time frames were saved from each simulated case.

In order to use the data in the proposed algorithm, we down-sampled each snapshot of our $512\times 512$ domain into four images of $256\times256$, and later used cubic interpolation \cite{suli2003introduction} to further reduce the resolution to $128\times128$.
Hence, from the $500$ microstructure evolution samples, we were able to generate 2,000 microstructure evolution samples of $128\times128$ resolution.
From this dataset, we have used 1,600 cases for training the DeepONet and 400 cases for testing the network accuracy.
Since the compositional field is randomly distributed spatially, the microstructure has no recognizable features at the first frame $t_0$.
The quick development of subdomains is then observed between frames $t_0$ and $t_{10}$, followed by a smooth and steady coalescence and growth of the microstructure from time frames $t_{10}$ to $t_{100}$.
We have trained our proposed model based on this observation starting at time frame, $t_{10}$, when the microstructure had reached a slow and steady development regime.

\subsection*{Training the autoencoder: Learning the latent microstructure representation}
\label{subsec: training autoencoder}

In this work, each microstructure evolution is represented by $(N_{\rm T}, N_{\rm x}, N_{\rm y}) = (80 \times 128 \times 128)$, with $N_{\rm T}$ representing the number of snapshots and $~N_{\rm x}\times~N_{\rm y}$ denoting the the spatial resolution along $x-$ and $y-$ direction, respectively.
To handle the entire feature space ($\mathbb{R}^{128 \times 128}$) 16,384 distinct features are required to represent the microstructure at each time step.
Subsequently, to compute the prediction for all 1,600 microstructure evolutions, we will have  $1,600 \times 80 \times 16,384 ~(\approx 2.5~\text{Billion})$ 32-bit floating data points. Learning microstructure dynamics from such a high-dimensional dataset is challenging. 

To circumvent issues pertaining to the data dimensionality and preparing the phase-field microstructure data for DeepONet training, we explored a couple of options.
First, we tried using Principal Component Analysis (PCA) with linear kernel, for reducing the dimensionality of the data \cite{brough2017extraction, niezgoda2013novel, gupta2015structure}.
The low-dimensional representation of the data obtained from PCA is a linear transformation of the high-dimensional data and discards the insignificant modes (eigen/singular) corresponding to the lower eigen/singular values ($\lambda_i$).
However, the system considered here is non-diffusive, which is confirmed by cumulative explained variance and energy distribution of the system over principal modes.
Therefore, using PCA for reducing the dimensionality of the microstructure description could result in the loss of valuable information, if only a convenient low number of principal components are considered.
A detailed explanation on PCA of microstructure dataset is presented in Supplementary Note 1.
Learning a non-linear mapping from a high-dimensional to a low-dimensional latent space is one way to compress data without losing as much information as in the PCA.
An autoencoder precisely does this by learning a non-linear transformation to a low dimensional latent space using an encoder.
The decoder learns the mapping to retrieve initial high-dimensional data from its latent representation.

In this study, we have used a convolutional autoencoder \cite{lee2020model} with convolutional layers in the encoder and transpose convolutional layers in the decoder as shown in Fig. \ref{fig: complete_algo}.
The encoder learns a nonlinear mapping of the high-dimensional microstructure data, $\phi (\bm x, \bm y, t)$, to a low-dimensional latent space represented by $\tilde{\bm{\phi}}(t)$ and is expressed as
\begin{subequations}
\begin{equation}\label{eq:encoder}
    \alpha_{\bm{\theta_{\text{enc}}}}: \bm{\phi}(\bm x,\bm y,t) \rightarrow \tilde{\bm{\phi}}(t),
\end{equation}
\begin{equation}\label{eq:decoder}
    \beta_{\bm{\theta_{\text{dec}}}}: \tilde{\bm{\phi}}(t) \rightarrow \hat{\bm{\phi}}(\bm x,\bm y,t), 
\end{equation}
\end{subequations}
where $\alpha$ and $\beta$ represent the mappings performed by the encoder and the decoder, respectively.
In Eq. \eqref{eq:encoder}, the encoder takes $\bm{\phi}(\bm x,\bm y,t) \in \mathbb{R}^{128 \times 128}$ as input, and maps it to $\tilde{\bm{\phi}}(t) \in \mathbb{R}^{l_{\rm d}}$, where $l_{\rm d}$ is the dimension of the latent space.
$\bm{\theta}_{\rm enc}$ represents the trainable parameters of the convolutional encoder.
Equation \eqref{eq:decoder} represents the decoder network, which takes the latent dimensional representation, $\tilde{\bm{\phi}}(t) \in \mathbb{R}^{l_{\rm d}}$ as the input and predicts the primitive microstructure, $\hat{\bm{\phi}}(\bm x,\bm y,t) \in \mathbb{R}^{128 \times 128}$, using transpose convolutional operations. The details of the autoencoder architecture are provided in \autoref{tab:conv_encoder_detail}.

$\bm{\theta}_{\text{ae}}$ represents the trainable parameters of the autoencoder.
These parameters are learned by minimizing the loss function, $\mathcal{L}_{\rm ae}$, which reads
\begin{equation}
    \mathcal{L}_{\rm ae} = \min_{\bm{\theta_{\rm ae}}=\{\bm{\theta}_{\text{enc}}, \bm{\theta}_{\text{dec}}\}} ||\bm{\phi}(x,y, t) - \hat{\bm{\phi}}(x,y,t; \bm{\theta}_{\rm ae})||_2^2.
\end{equation}

Alternatively, the autoencoder provides a low-dimensional representation of the microstructure by learning a non-linear transformation to a latent space with $l_{\rm d}$ features.
We also observe that it is easier to learn the microstructure dynamics in the latent space representation, learned by the autoencoder, than the original primitive form of the microstructure in real space.
This is due to the presence of several high gradient region in the original form of the microstructure as shown in Fig. \ref{fig: distributions}(a).
These high gradient regions in the solution are due to the nature of the governing Cahn-Hilliard equation.
The latent microstructure representation in Fig. \ref{fig: distributions}(b) is smoother and does not have high gradient regions.
The latent representation of the data offers higher regularity and therefore, we achieve faster convergence during the training of the surrogate neural network model.

\subsection*{Training the DeepONet: Learning the microstructure dynamics in lower dimensions}
\label{subsec:trainingDeepONet}

Neural operators generate nonlinear mappings across infinite dimensional function spaces on bounded domains, giving a novel simulation framework for multidimensional complex dynamics prediction in real time.
Once properly trained, such models are discretization invariant, which means they share the same network parameters regardless of how the underlying functional data is parameterized.
DeepONet, originally proposed by Lu and coworkers~\cite{lu2021learning}, allows the mapping between infinite dimensional functions using deep neural networks.
This subsection provides a detailed description of the training of DeepONet to model the evolution of the microstructure in the latent dimension.

\begin{figure}[H]
    \centerline{\includegraphics[width=\textwidth]{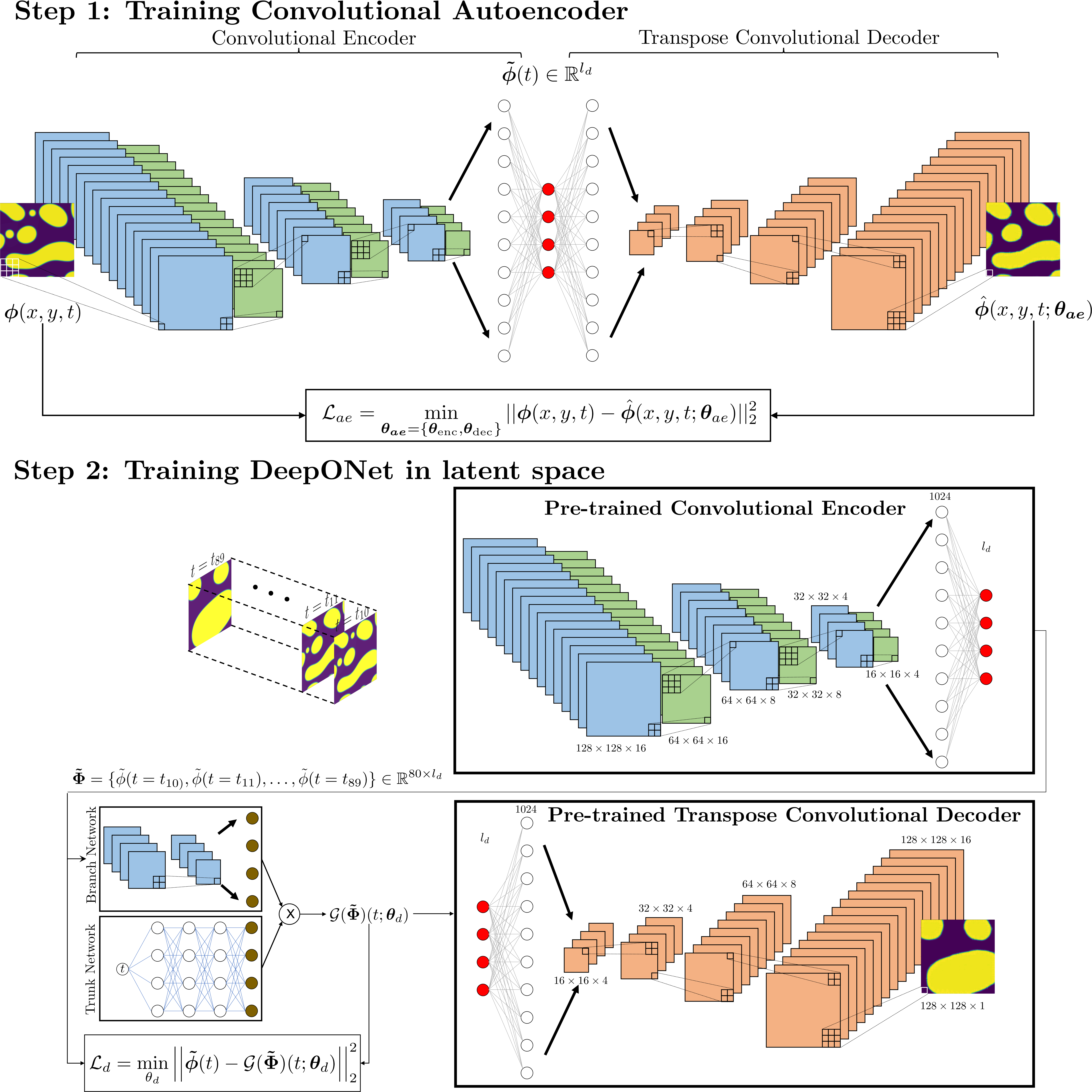}}
    \caption{{\bf Schematic representation of DeepONet with convolutional autoencoder.} Step 1 involves training of the convolutional autoencoder to minimize $\mathcal{L}_{\rm ae}$. The encoder learns a suitable transformation from the high-dimensional microstructure to a low-dimensional latent space through a series of convolution (blue layers) and MaxPooling (green layers) operations. The decoder remaps the latent representation of the microstructure back to the original, real space by performing transpose convolution (orange layers) operations. Detailed description of the architecture is provided in \autoref{tab:conv_encoder_detail}. In step 2, we train the DeepONet in the latent space to minimize $\mathcal{L}_{\rm d}$. The entire history of 80 steps is encoded by the pre-trained convolutional encoder as $\tilde{\Phi}$. DeepONet learns to predict $\tilde{\phi}(t)$ at any desired time $t$, fed to the trunk network. The latent representation of the microstructure predicted by DeepONet is then re-mapped back to the primitive space by the transpose convolutional decoder. }
    \label{fig: complete_algo}
\end{figure}

The unstacked DeepONet architecture is made up of two concurrent deep neural networks:
one encodes the input function at fixed sensor locations (branch network),
while the other represents the domain of the output function (trunk network).
Time, $t \in \mathbb{R}^{1}$, is given as input to the trunk network while $\tilde{\bm{\Phi}} = \{\tilde{\phi}(t_{10}), \tilde{\phi}(t_{11}), \ldots, \tilde{\phi}(t_{89})\} \in \mathbb{R}^{80\times l_{\rm d}}$ is the input fed to the branch network.
$\tilde{\bm{\Phi}}$ represents the phase field in the latent dimension, $l_{\rm d}$, for all the 80 time steps available in the given dataset.
The goal of the DeepONet is to learn the solution operator, $ \tilde{\phi}(t) \approx \hat{\tilde{\phi}}(t) = \mathcal{G}(\tilde{\bm{\Phi}})(t)$ from the 1,600 microstructure evolutions provided in the training dataset.
The output of the DeepONet is a vector $\in \mathbb{R}^{l_{\rm d}}$ and is expressed as $\mathcal{G}(\tilde{\bm{\Phi}})(t; \bm\theta_{\rm d})$, where $\bm{\theta}_{\rm d} = \left\{\mathbf W_{\rm d}, \mathbf b_{\rm d} \right\}$ includes the trainable weights, $\mathbf W_{\rm d}$, and biases, $\mathbf b_{\rm d}$, of the DeepONet model.
The framework of the DeepONet allows the branch network to have a flexible architecture.
To model the microstructure evolution, we have considered a fully-connected neural network for the trunk network.
Due to the high-dimensional nature of the branch network input, $\mathbb{R}^{80 \times l_{\rm d}}$, a convolutional neural network is used as the branch network because it utilizes the same kernels across time axis and enables the branch network to encode the entire history in a memory efficient manner.
Hence, the input has to be reshaped to $\mathbb{R}^{80 \times \sqrt{l_{\rm d}} \times \sqrt{l_{\rm d}}}$ before feeding it to the branch network.
The network architecture is presented in Fig. \ref{fig: complete_algo}.
The trainable parameters of the DeepONet, $\bm{\theta}_{\rm d}$, are obtained by minimizing a loss function, $\mathcal l_{\rm d}$, defined as: 
\begin{equation}\label{eq:lossfunction_deeponet}
    \mathcal{L}_{\rm d} = \min_{\theta_{\rm d}} \left| \left| \tilde{\bm{\phi}}(t) - \mathcal{G}(\tilde{\bm{\Phi}})(t; \bm\theta_{\rm d}) \right| \right|_2^2 ,
\end{equation}
where $\tilde{\phi}(t)$ is the ground truth for the low-dimensional phase field representation at time,
$t$ obtained from the convolutional encoder.
The trained DeepONet is used to predict $\tilde\phi(t) \in \mathbb{R}^{l_{\rm d}}$.
The output of the DeepONet is fed into the transposed convolutional decoder to predict, $\hat{\phi}(t) \in \mathbb{R}^{128\times128}$.
The DeepONet is trained using the Adam optimizer \cite{kingma2014adam}.
The implementation has been carried out using the \texttt{TensorFlow} framework \cite{abadi2015tensorflow}.
We use Xavier Initialization \cite{glorot2010understanding} to initialize the weights of all the models.

\subsection*{Error metrics}

 The $L^2$ norm of relative error, $\mathcal{D}$, is used as the evaluation metric to analyze the performance of each model considered in this study. $\mathcal{D}$ is defined as:

\begin{equation}
    \label{eqn: l2 norm}
    \mathcal{D} = \frac{\sum_n \sum_x \sum_y \sum_t \left( \phi^{(n)}(x,y,t) - \hat{\phi}^{(n)}(x,y,t; \bm{\theta}) \right)^2}{\sum_n \sum_x \sum_y \sum_t \phi^{(n)}(x,y,t)^2},
\end{equation}
where $n$ corresponds to the $n^{th}$ sample of the given dataset.


\section*{DATA AVAILABILITY}
\par{
The data that support the findings of this study are available from the corresponding authors upon reasonable request.
}


\section*{CODE AVAILABILITY}
The codes used to calculate the results of this study are available from the corresponding author upon reasonable request.


\section*{ACKNOWLEDGEMENTS}
R.D. and G.E.K. acknowledge funding under the \textit{Beyond}Fingerprinting Sandia Grand Challenge Laboratory Directed Research and Development (GC LDRD) program.
V.O. and K.S. acknowledge Dr. Elton Chen from Sandia National Laboratories for their valuable discussions that enabled us to better understand the MEMPHIS code.
V.O. and K.S. acknowledge Dr. Aniruddha Bora for his insightful suggestions and valuable discussions at various stages of this project.
The phase-field framework is supported by the Center for Integrated Nanotechnologies (CINT), an Office of Science user facility operated for the U.S. Department of Energy. This research was conducted using computational resources and services at the Center for Computation and Visualization, Brown University.
Sandia National Laboratories is a multi-mission laboratory managed and operated by National Technology and Engineering Solutions of Sandia, LLC., a wholly owned subsidiary of Honeywell International, Inc., for the U.S.\ Department of Energy National Nuclear Security Administration under contract DE-NA0003525. This paper describes objective technical results and analysis. Any subjective views or opinions that might be expressed in the paper do not necessarily represent the views of the U.S.\ Department of Energy or the United States Government.   


\section*{AUTHOR CONTRIBUTIONS}

R.D. and G.E.K. designed the study and supervised the project. V.O. and K.S. developed the autoencoder--DeepONet architecture and the hybrid model. V.O., K.S. and S.G. analyzed the results. V.O., K.S., S.G., R.D. and G.E.K wrote the manuscript.  


\section*{COMPETING INTERESTS}
The authors declare no competing interests.


\section*{ADDITIONAL INFORMATION}
{\bf Supplementary information} is available for this paper.


\bibliography{NPJCOMPUMATS-01941-R2}
\clearpage


\section*{FIGURES}

\section*{TABLES}
\begin{table}[!htb]
    \centering
    \caption{Details of the hyper-parameters used in the convolutional autoencoder.}
    \label{tab:conv_encoder_detail}
    \begin{tabular}{llcccccc}
    \hline
    \hline
     & Layer & Kernel Size & Width & Activation & Output\\ \hline \hline
     1 & Conv2D & $3\times3$ & $16$ & ReLU & $128\times128\times16$\\
     2 & Max-Pool & $2\times2$ &  &  & $64\times64\times16$\\
     3 & Conv2D & $3\times3$ & $8$ & ReLU & $64\times64\times8$\\
     4 & Max-Pool & $2\times2$ &  &  & $32\times32\times8$\\
     5 & Conv2D & $3\times3$ & $4$ & ReLU & $32\times32\times4$\\
     6 & Max-Pool & $2\times2$ &  & & Reshaped to $1024$\\
     \hline \\
     7 & Fully connected & & $l_{\rm d}$ & Linear &$l_{\rm d}$ \\
     8 & Fully connected & & $1024$ & ReLU & Reshaped to $16 \times 16 \times 4$\\ \\
     \hline
     9 & Transpose Conv2D & $2\times2$ &  & ReLU & $32\times32\times4$\\
     10 & Transpose Conv2D & $2\times2$ &  & ReLU & $64\times64\times8$\\
     11 & Transpose Conv2D & $2\times2$ &  & ReLU & $128\times128\times16$\\
     12 & Transpose Conv2D & $3\times3$ &  & ReLU & $128\times128\times1$\\
    \hline
        & \small{$\bm{l_{\rm d}}$ \textbf{is the dimension of latent space}}
    \end{tabular}
\end{table}

~\clearpage
\begin{table}[H]
\caption{Detailed survey of different latent dimension size, $l_{\rm d}$, network architecture, and non-linear activation functions.}
\label{table: latent_dim}
\begin{center}
\begin{tabular}{cccccc}
\hline
\hline
$l_{\rm d}$          & \textbf{\begin{tabular}[c]{@{}l@{}}DeepONet \\ Architecture\end{tabular}} & \textbf{Activation} & $\mathcal{D}_{train}$ & $\mathcal{D}_{\rm test}$ & \textbf{\begin{tabular}[c]{@{}l@{}}DeepONet training \\  time per 1,000 \\ epochs (s) \end{tabular}} \\
\hline
\hline \vspace{-1mm}

                     &                                                                           &                     &                &               &   
                     \\
\multirow{3}{*}{196} & \multirow{3}{*}{Architecture 1}                                           & ReLU                & 0.03621        & 0.06803       & 56                                                                                                  \\
                     &                                                                           & $\tanh$                & 0.02177        & 0.06233       & 56                                                        \                                         \\
                     &                                                                           & $\sin$                 & 0.01408        & 0.01620      & 57                                                                                                 \\ \\
                    \hline \vspace{-1mm}
                    \\
\multirow{3}{*}{100} & \multirow{3}{*}{Architecture 2}                                           & ReLU                & 0.04076        & 0.05991       & 31                                                                                                  \\
                     &                                                                           & $\tanh$                & 0.03196        & 0.04699       & 30                                                                                                  \\
                     &                                                                           & $\sin$                 & 0.02684        & 0.03679       & 31                                                                                                  \\ \\
\hline
\\
\multirow{3}{*}{64}  & \multirow{3}{*}{Architecture 3}                                           & ReLU                & 0.06708        & 0.07791       & 31                                                                                                  \\
                     &                                                                           & $\tanh$                & 0.04773        & 0.06013       & 35                                                                                                  \\
                     &                                                                           & $\sin$                 & 0.04781        & 0.05739       & 32                                                                                                  \\ \\
                     \hline \vspace{-1mm}
                     \\
\multirow{3}{*}{25}  & \multirow{3}{*}{Architecture 4}                                           & ReLU                & 0.16527        & 0.20097       & 19                                                                                                  \\
                     &                                                                           & $\tanh$                & 0.16507        & 0.20134       & 19                                                                                                  \\
                     &                                                                           & $\sin$                 & 0.16551        & 0.20167       & 20                                                                                                  \\ \\
\hline \vspace{-1mm}
\\
\multirow{3}{*}{9}   & \multirow{3}{*}{Architecture 5}                                           & ReLU                & 0.31536        & 0.3186        & 10                                                                                                  \\
                     &                                                                           & tanh                & 0.31523        & 0.31903       & 11                                                                                                  \\
                     &                                                                           & sin                 & 0.31539        & 0.31876       & 11 \\
                    \\ \hline \vspace{-1mm}
\end{tabular}

\vspace{1mm}

\begin{tabular}{ccc}
\hline
\hline
\textbf{Architecture} & \textbf{Branch Network}              & \textbf{Trunk Network}   \\
\hline
\hline \vspace{1mm}
1                     & 3$\times${[}conv(32,(3,3)){]} + {[}1960{]} & 2$\times${[}100{]} + {[}1960{]} \\ \vspace{1mm}
2                     & 2$\times${[}conv(32,(3,3)){]} + {[}1100{]} & 2$\times${[}100{]} + {[}1100{]} \\ \vspace{1mm}
3                     & 2$\times${[}conv(32,(3,3)){]} + {[}512{]} & 2$\times${[}100{]} + {[}512{]} \\ \vspace{1mm}
4                     & 1$\times${[}conv(64,(3,3)){]} + {[}500{]}  & 2$\times${[}100{]} + {[}500{]}  \\ \vspace{1mm}
5                     & 1$\times${[}conv(128,(3,3)){]} + {[}180{]}  & 2$\times${[}100{]} + {[}180{]} \\ 
\hline
\end{tabular}

\end{center}
\end{table}


    


\begin{table}[H]
    \centering
    \caption{A comparison between computational time for high-fidelity phase-field simulations (MEMPHIS) and proposed hybrid model (Hybrid) for a single microstructure evolution realization from time frame $t_1$ to time frame $t_{105}$.}
    \begin{tabular}{lll}
        \\
        \hline
        \hline
                                & \multicolumn{2}{c}{\textbf{Computational time}} \\
        \textbf{Time frames}    & \textbf{MEMPHIS}        & \textbf{Hybrid}       \\
        \hline
        \hline
        $t_{1}$ to $t_{30}$     & 135 mins                & 135 mins               \\
        $t_{30}$ to $t_{40}$    & 45 mins                 & 2 sec                 \\
        $t_{40}$ to $t_{60}$    & 90 mins                 & 90 mins                \\
        $t_{60}$ to $t_{70}$    & 45 mins                 & 2 sec                 \\
        $t_{70}$ to $t_{90}$    & 90 mins                 & 90 mins                \\
        $t_{90}$ to $t_{100}$   & 45 mins                 & 2 sec                 \\
        $t_{100}$ to $t_{105}$  & 23 mins                 & 23 mins                \\
                                &                         &                       \\
        \hline
        \textbf{Total time}     & 473 mins                & 338.1 mins            \\
        \hline
    \end{tabular}
    \label{table: integration}
    \end{table}


~\clearpage
\section*{Algorithms}

\begin{algorithm}
\caption{Integration of DeepONet with high-fidelity phase field simulator (MEMPHIS)}
\label{alg: mod}
    \begin{algorithmic}
        \Require $\phi_0:$ Initial condition:  $\phi(x, y, 0)$
        \Require $N_T:$ Number of total time steps
        \Require $n_t$: Initial number of time steps to be simulated by MEMPHIS 
        \Require ${\rm DON}_{nt}$: Number of time steps to be leaped by DeepONet 
        \State $n \gets 0$
        \Comment{Initialize}
        \While{$n~!=~N_T$}
            \State $\phi_{n_t} \gets {\rm MEMPHIS}(\phi_{0}, n_t)$
            \Comment{Solution from MEMPHIS}
           \State $\phi_{n_t + {\rm DON}_{nt}} \gets {\rm DeepONet}(\phi_{n_t})$ \Comment{Prediction from DeepONet}
           \State $\phi_{0} \gets \phi_{n_t + {\rm DON}_{nt}}$
           \Comment{Update the input for MEMPHIS}
            \State $n \gets {n_t + {\rm DON}_{nt}}$ \Comment{Leaping $n$: MEMPHIS + DeepONet} 
        \EndWhile
    \end{algorithmic}
\end{algorithm}

\clearpage

\section*{Supplementary Note 1: Principal Component Analysis (PCA) of microstructure dataset}

\begin{figure}[H]\renewcommand\figurename{Supplementary Figure} 
    \centerline{\includegraphics[width=\textwidth]{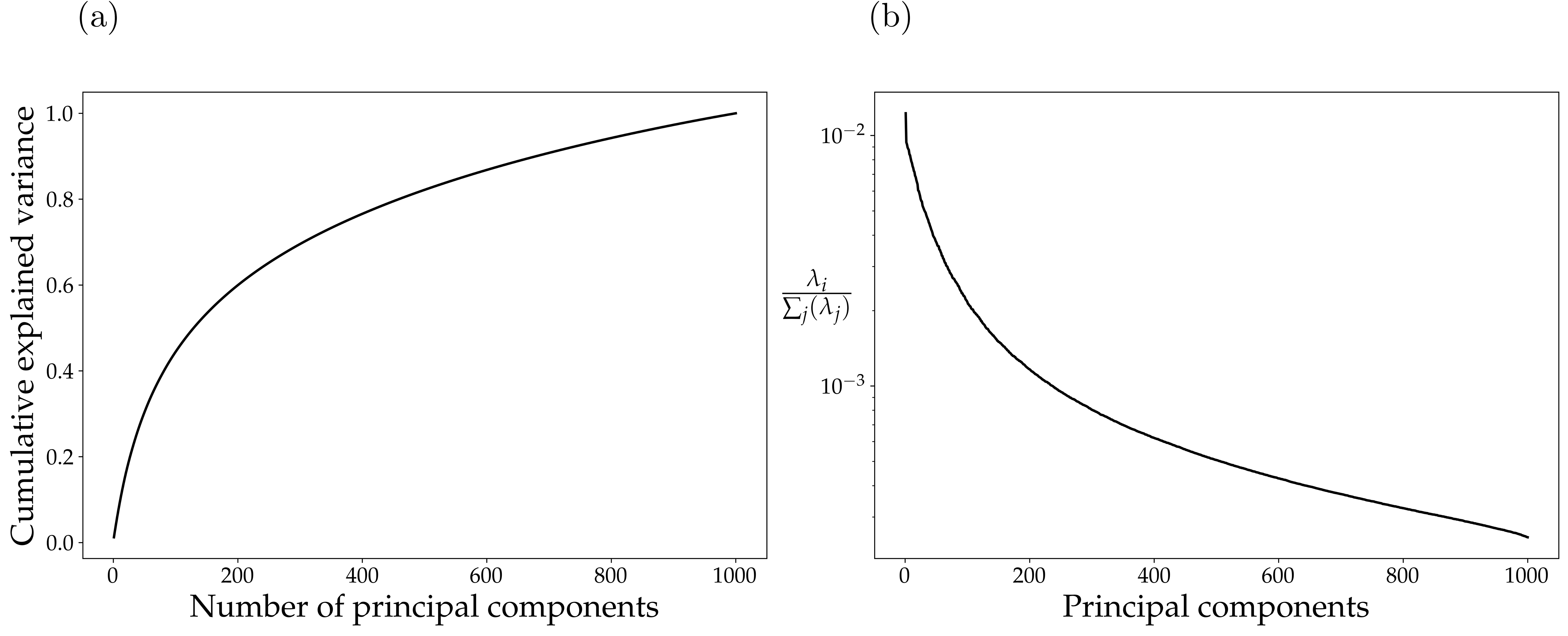}}
    \caption{(a) The cumulative explained variance curve on the dominant 1,000 eigenvalues of the phase-field microstructure (b) The energy curve on the dominant 1,000 eigenvalues of the phase-field microstructure. The two plots suggest that microstructure data represent a non-diffusive system. }
    \label{fig: PCA results}
\end{figure}

To circumvent issues pertaining to the data dimensionality and preparing the phase-field microstructure data for DeepONet training, we explored a couple of dimensionality reduction options.
First, we tried using Principal Component Analysis (PCA) with linear kernel, for reducing the dimensionality of the data \cite{brough2017extraction, niezgoda2013novel, gupta2015structure}.
The low-dimensional representation of the data obtained from PCA is a linear transformation of the high-dimensional data and discards the insignificant modes (eigen/singular) corresponding to the lower eigen/singular values ($\lambda_i$).
However, the system considered here is not diffusive and this is shown in Suppl. Fig.~\ref{fig: PCA results}.
Therefore, using PCA for reducing the dimensionality of the microstructure description could result in the loss of valuable information, if only a convenient low number of principal components are considered. Learning a non-linear mapping from a high dimensional to a low-dimensional latent space is one way to compress data without losing as much information as in the PCA.

\clearpage

\section*{Supplementary Note 2: Experiment on number of training samples required}

\begin{figure}[H]
	\centering
		\begin{tikzpicture}
		\begin{axis}[
		legend cell align=left,
		width=.6\textwidth,
		xlabel={Time step $(t)$},
		ylabel={\large{$\mathcal{D}_{\rm test}(t)$}}, 
		xmin=88, xmax=101,
		ymin=0, ymax=3e-1,
		legend pos=north west, legend cell align=left, legend style={font=\tiny},
		xmajorgrids=true,
		ymajorgrids=true,
		grid style=dashed,
		] 	
		\addplot[color=blue,mark=*,semithick, mark options={fill=markercolor}]
		coordinates{(90,0.2062785) (91,0.20960072) (92,0.21300773) (93,0.2165105) (94,0.22009455) (95,0.223863) (96,0.22768939) (97,0.23136124) (98,0.23477928) (99,0.23828425) };
		
		\addplot[color=red,mark=square*,semithick, mark options={fill=markercolor}]
		coordinates{(90,0.09350135) (91,0.09660445) (92,0.09985614) (93,0.10337224) (94,0.107212946) (95,0.111268155) (96,0.11556207) (97,0.119776756) (98,0.12391999) (99,0.12836936) };
		
		 \addplot[color=black,mark=triangle*,semithick, mark options={fill=markercolor}]
		coordinates{(90,0.041694824) (91,0.043805156) (92,0.045998372) (93,0.048448183) (94,0.051251024) (95,0.054360725) (96,0.057903152) (97,0.061578255) (98,0.065516666) (99,0.07014392) };
		
		\addplot[color=magenta,mark=diamond*,semithick, mark options={fill=markercolor}]
		coordinates{(90,0.018845389) (91,0.02112368) (92,0.023655823) (93,0.026688272) (94,0.03034132) (95,0.034539163) (96,0.039455354) (97,0.04482879) (98,0.05083209) (99,0.057964493)};

		\legend{400 samples, 800 samples, {1,200} samples, {1,600} samples}
		\end{axis}
		\end{tikzpicture}
	\caption{Sensitivity of the proposed surrogate model to the number samples in training data. $\mathcal{D}_{\rm test}(t)$ represents the relative squared error computed across the samples in test dataset at different time steps. We analyze the performance of each model on the basis of $\mathcal{D}_{\rm test}(t)$ at forecasting time steps. }
	\label{fig: train_samples}
\end{figure}
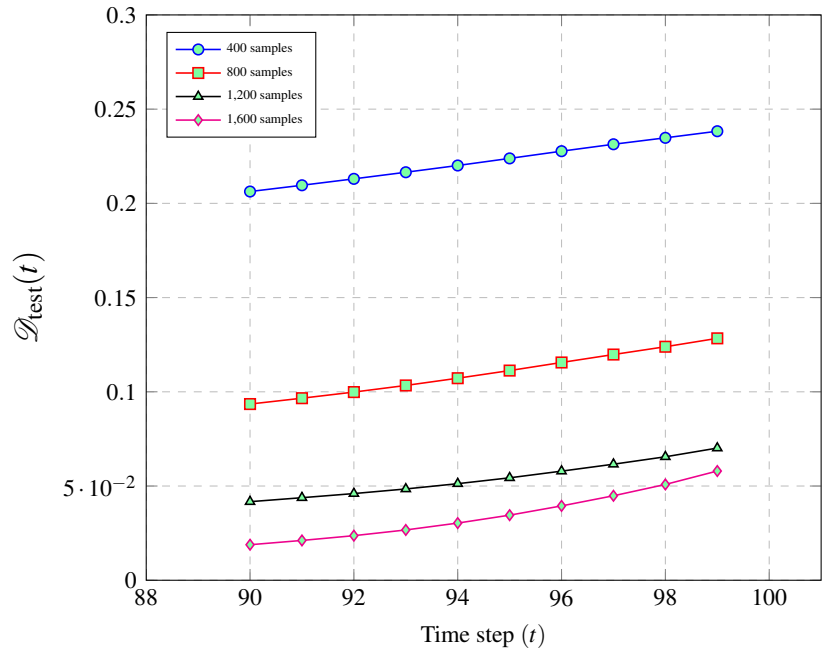

We carried out a computational experiment to analyze the sensitivity of the proposed approach to the number of samples used for training. We consider training datasets with 25\%, 50\%, 75\% and all the 1,600 training samples. We trained separate Autoencoder-DeepONet models on each of the four datasets. The model performance is evaluated on the basis of forecasting errors on test data, shown in Suppl. Fig.~\ref{fig: train_samples}. We observe better accuracy in model's predictions while increasing the number of training samples. The model trained with 1,200 data samples shows similar accuracy as the best model trained with 1,600 data samples.

\clearpage

\section*{Supplementary Note 3: Experiment on choice of training loss function}

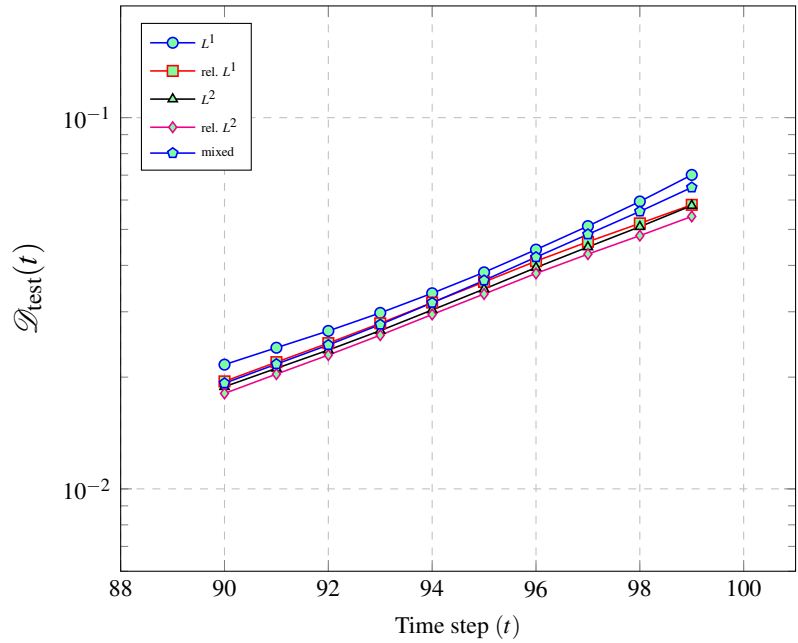
\begin{figure}[H]
	\centering
		\begin{tikzpicture}
		\begin{semilogyaxis}[
		legend cell align=left,
		width=.6\textwidth,
		xlabel={Time step $(t)$},
		ylabel={\large{$\mathcal{D}_{\rm test}(t)$}}, 
		xmin=88, xmax=101,
		ymin=6e-03, ymax=2e-1,
		legend pos=north west, legend cell align=left, legend style={font=\tiny},
		xmajorgrids=true,
		ymajorgrids=true,
		grid style=dashed,
		] 	
		\addplot[color=blue,mark=*,semithick, mark options={fill=markercolor}]
		coordinates{(90,0.021613017) (91,0.024002312) (92,0.026639055) (93,0.029791014) (94,0.033654634) (95,0.038312055) (96,0.044108186) (97,0.05103378) (98,0.05945987) (99,0.07011384) };
		
		\addplot[color=red,mark=square*,semithick, mark options={fill=markercolor}]
		coordinates{(90,0.019478532) (91,0.021949572) (92,0.024695328) (93,0.027933223) (94,0.031761076) (95,0.03609584) (96,0.041055523) (97,0.046303857) (98,0.051906228) (99,0.058239322) };
		
		 \addplot[color=black,mark=triangle*,semithick, mark options={fill=markercolor}]
		coordinates{(90,0.018845389) (91,0.02112368) (92,0.023655823) (93,0.026688272) (94,0.03034132) (95,0.034539163) (96,0.039455354) (97,0.04482879) (98,0.05083209) (99,0.057964493) };
		
		\addplot[color=magenta,mark=diamond*,semithick, mark options={fill=markercolor}]
		coordinates{(90,0.018077746) (91,0.020379711) (92,0.02293637) (93,0.025942672) (94,0.029499227) (95,0.033492398) (96,0.038059305) (97,0.042885307) (98,0.048101045) (99,0.054191492) };
		
		\addplot[color=blue,mark=pentagon*,semithick, mark options={fill=markercolor}]
		coordinates{(90,0.01925975) (91,0.021683708) (92,0.024413647) (93,0.027714115) (94,0.031726338) (95,0.036443837) (96,0.042093586) (97,0.04849231) (98,0.055893935) (99,0.06487371)};
		
		\legend{$L^1$, rel. $L^1$, $L^2$, rel. $L^2$, mixed}
		\end{semilogyaxis}
		\end{tikzpicture}
	\caption{Variation of forecasting error on test data for models with different autoencoder loss functions, $\mathcal{L}_{\rm ae}$, specifically $L^1$, relative $L^1$, $L^2$, relative $L^2$ and mixed loss($L_2$ loss for the initial 5,000 epochs and $L^1$ loss for the remaining epochs). $\mathcal{D}_{\rm test}(t)$ represents the relative $L^2$ error computed across the samples in test dataset at different time steps. }
	\label{fig: AE_losses}
\end{figure}

We investigate the effect of using different loss functions on training of the autoencoder models. Specifically, we trained autoencoders by minimizing (1) $L^1$ loss, (2) relative $L^1$ loss, (3) $L^2$ loss, (4) relative $L^2$ loss, and (5) mixed loss ($L^2$ loss for the initial 5,000 epochs and $L^1$ loss for the remaining epochs). Now, (i) we train DeepONet model with architecture 1 on each of the learned latent microstructure data and (ii) re-transform the DeepONet predictions using pre-trained decoder to retrieve the microstructure. Finally, we analyze the model performance by computing the forecasting error, $\mathcal{D}_{\rm test}(t)$, on unseen test data, shown in Suppl. Fig.~\ref{fig: AE_losses}. We observe that, models whose autoencoder is trained on $L^2$ loss is better than $L^1$ loss. Similarly, mean values of relative $L^1$ and $L^2$ are a better choice for autoencoder loss, $\mathcal{L}_{\rm ae}$, than the mean of $L^1$ and $L^2$, respectively. In general, all the models perform well consistently.

\clearpage


\section*{Supplementary Note 4: Error in forecasting}

\begin{figure}[H]
	\centering
	\subfloat[Forecasting error on train dataset.]{
		\begin{tikzpicture}
		\begin{semilogyaxis}[
		legend cell align=left,
		width=.475\textwidth,
		xlabel={Time step ($t$)},
		ylabel={$\mathcal{D}_{\rm train}(t)$}, 
		xmin=88, xmax=101,
		ymin=1e-3, ymax=0.8,
		legend pos=south east, legend cell align=left, legend style={font=\tiny},
		xmajorgrids=true,
		ymajorgrids=true,
		grid style=dashed,
		] 	
		\addplot[color=blue,mark=*,semithick, mark options={fill=markercolor}]
		coordinates{(90, 0.30228668) (91, 0.30256444) (92, 0.30298913) (93, 0.30354163) (94, 0.30422303) (95, 0.30505812) (96,0.30605567) (97, 0.30723882) (98, 0.3085818) (99,0.3100722)};
		\addplot[color=red,mark=square*,semithick, mark options={fill=markercolor}]
		coordinates{(90, 0.13275073)(91, 0.13366967)(92, 0.13477965)(93, 0.13614057)(94, 0.1377609)(95, 0.13960434)(96, 0.14175546)(97, 0.14424056)(98, 0.14705104)(99, 0.15018836)};
	    \addplot[color=black,mark=triangle*,semithick, mark options={fill=markercolor}]
		coordinates{(90, 0.03252058)(91, 0.034327194)(92, 0.03652512)(93, 0.039270904)(94, 0.04249952)(95, 0.046075355)(96, 0.050114464)(97, 0.05456493)(98, 0.059383273) (99, 0.064500496)};
		\addplot[color=magenta,mark=diamond*,semithick, mark options={fill=markercolor}]
		coordinates{(90, 0.019845879)(91, 0.021879444)(92, 0.0240936)(93, 0.026680293)(94, 0.02958439)(95, 0.032683406)(96, 0.036126778)(97, 0.03988244)(98, 0.04392311)(99, 0.048263017)};
		\addplot[color=blue,mark=pentagon*,semithick, mark options={fill=markercolor}]
		coordinates{(90, 0.01653671)(91, 0.019534444)(92, 0.022646835)(93, 0.026062855)(94, 0.029728651)(95, 0.033495497)(96, 0.037511945)(97, 0.041720342)(98, 0.04607762)(99, 0.050582122)
};      
        \legend{$l_{\rm d}=9$,$l_{\rm d}=25$,$l_{\rm d}=64$, $l_{\rm d}=100$,$l_{\rm d}=196$}

		\end{semilogyaxis}
		\end{tikzpicture}
	}
	\subfloat[Forecasting error on test dataset.]{
		\begin{tikzpicture}
		\begin{semilogyaxis}[
		legend cell align=left,
		width=.475\textwidth,
		xlabel={Time step ($t$)},
		ylabel={$\mathcal{D}_{\rm test}(t)$}, 
		xmin=88, xmax=101,
		ymin=1e-3, ymax=0.8,
		legend pos=south east, legend cell align=left, legend style={font=\tiny},
		xmajorgrids=true,
		ymajorgrids=true,
		grid style=dashed,
		] 	
		\addplot[color=blue,mark=*,semithick, mark options={fill=markercolor}]
		coordinates{(90, 0.35101995)(91, 0.3508237)(92, 0.35069647)(93, 0.35062924)(94, 0.35063782)(95, 0.3507673)(96, 0.35101008)(97, 0.3513878)(98, 0.35191512)(99, 0.35261008)};
		\addplot[color=red,mark=square*,semithick, mark options={fill=markercolor}]
		coordinates{(90, 0.19623238)(91, 0.19693744)(92, 0.19782266)(93, 0.19896963)(94, 0.20033993)(95, 0.20187992)(96, 0.20374463)(97, 0.20579368)(98, 0.2080668)(99, 0.2106768)};
		 \addplot[color=black,mark=triangle*,semithick, mark options={fill=markercolor}]
		coordinates{(90, 0.05460544)(91, 0.05646596)(92, 0.058791008)(93, 0.061570086)(94, 0.06480004)(95, 0.068406284) (96, 0.07246108)(97, 0.076688394) (98, 0.08115602) (99, 0.08589316)};
		\addplot[color=magenta,mark=diamond*,semithick, mark options={fill=markercolor}]
		coordinates{(90, 0.040148288)(91, 0.04196502)(92, 0.04401828)(93, 0.04636412)(94, 0.049014565)(95, 0.051728066)(96, 0.054780617)(97, 0.057906378)(98, 0.06126587)(99, 0.06484309)};
		\addplot[color=blue,mark=pentagon*,semithick, mark options={fill=markercolor}]
		coordinates{(90, 0.041245773)(91, 0.04394478)(92, 0.046905804)(93, 0.0501773)(94, 0.0537687)(95, 0.05742191)(96, 0.061363835)(97, 0.06531732)(98, 0.069406636)(99, 0.0736145)};
		\legend{$l_{\rm d}=9$,$l_{\rm d}=25$,$l_{\rm d}=64$,$l_{\rm d}=100$,$l_{\rm d}=196$}
		\end{semilogyaxis}
		\end{tikzpicture}
	}\\
	
	\caption{(a) and (b) represents the relative $L^2$ error for predictions on the train and the test dataset. $\mathcal{D}_{\rm train}(t)$, $\mathcal{D}_{\rm test}(t)$ represents the mean of relative $L^2$ error computed on train and test datasets respectively. }
	\label{fig: forecasting_error}
\end{figure}
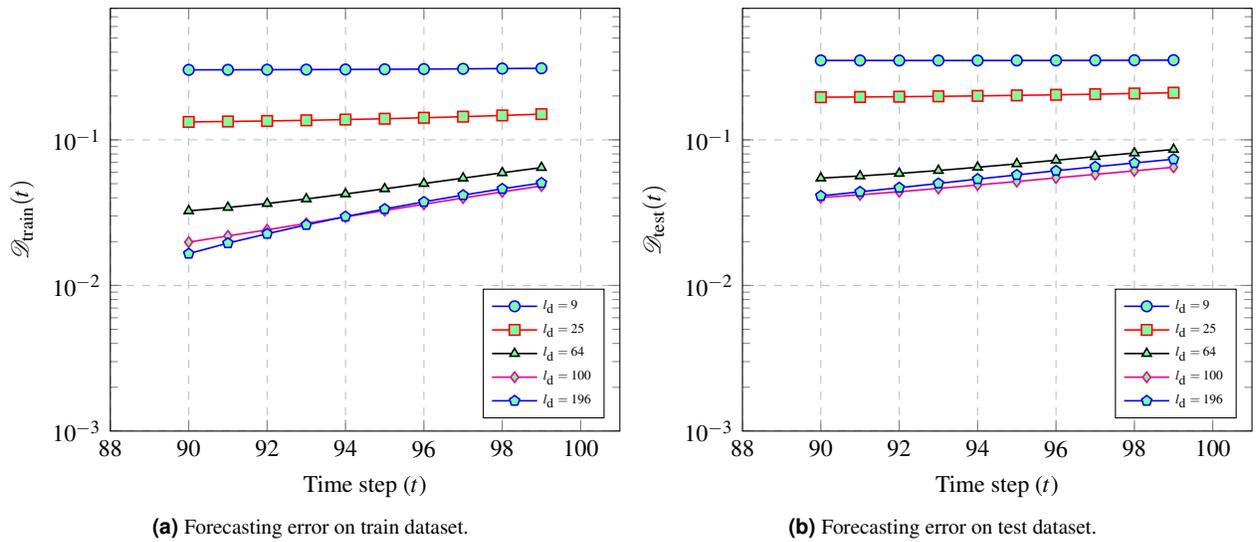

A closer look at the forecasting predictions offers further insights on the DeepONet predictive performance. We computed the mean of the relative $L^2$ error, $\mathcal{D}(t)$ for forecasting time frames $t = \{t_{90}, t_{91}, ..., t_{99}\}$, across the training and the testing datasets for all the latent dimensions considered in this study and plotted these values in Suppl. Fig.~\ref{fig: forecasting_error}. We observe that the mean relative $L^2$ error reduces by increasing the latent dimension of the autoencoder model. In other words, the model with larger latent spaces is able to predict microstructures better while forecasting. This is intuitive because a larger dimension of the latent space implies that there are more basis functions to express the encoded information about the microstructure and its evolution, and therefore the network has an improved representation capability. However, this trend seems to saturate beyond $l_{\rm d} = 100$. For the model with $l_{\rm d} = 100,\;196$, the forecasting error is always less than $6\%$. The logarithm of the relative $L^2-$ error linearly increases for $l_{\rm d}=64,\;100$ and $196$ for forecasting time step, whereas for $l_{\rm d}=9$ and $l_{\rm d}=25$, the error is high and remains constant.

\clearpage


\section*{Supplementary Note 5: Sensitivity to noise}

We investigate how the accuracy of prediction varies by systematically increasing the noise levels in the model input. For this analysis, we considered the best model with $l_d=196$ and DeepONet architecture 1 with $\sin$ activation functions. We added a Gaussian white noise with mean zero and standard deviations, $\sigma = 0.5\%, 1\%, 2\%, 3\%, 4\%, 5\%, 10\%$ to the input. To evaluate the model performance, we used the relative $L^2$ norm, $\mathcal{D}$, computed across all the time steps. $\mathcal{D}$ is calculated across the samples present in the test dataset. 

\begin{figure}[H]
	\centering
		\begin{tikzpicture}
		\begin{semilogyaxis}[
		legend cell align=left,
		width=.6\textwidth,
		xlabel={\% Noise levels},
		ylabel={\large{$\mathcal{D}_{\rm test}$}}, 
		xmin=0, xmax=10.8,
		ymin=1e-02, ymax=1e-1,
		legend pos=south east, legend cell align=left, legend style={font=\tiny},
		xmajorgrids=true,
		ymajorgrids=true,
		grid style=dashed,
		] 	
		\addplot[color=blue,mark=*,semithick, mark options={fill=markercolor}]
		coordinates{(0.5, 0.04344) (1, 0.04314) (2, 0.04260) (3, 0.04282) (4, 0.04300) (5, 0.04364) (10,0.04738) };
		\end{semilogyaxis}
		\end{tikzpicture}
	\caption{Sensitivity of the proposed surrogate model to different levels of noise added to the test data. $\mathcal{D}_{\rm test}$: represents the mean of relative $L^2$ error computed on test dataset for all time steps. }
	\label{fig: sensitivity_to_noise}
\end{figure}
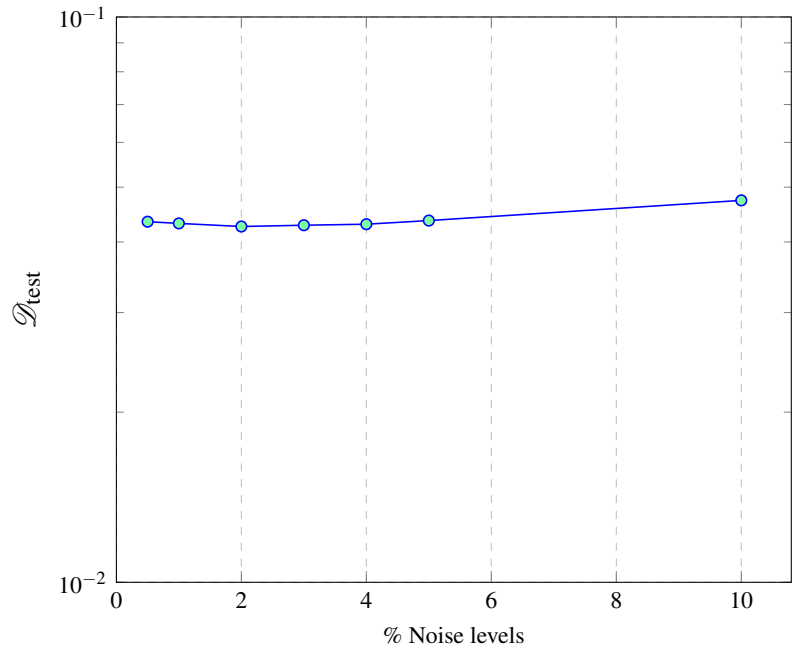

From Suppl. Table~\ref{table: noise} and Suppl. Fig.~\ref{fig: sensitivity_to_noise}, the relative $L^2$ norm does not increase much when noise is added to the model input. Previous studies \cite{vincent2008extracting, vincent2010stacked, gondara2016medical} illustrated the capability of the autoencoders to denoise noisy images. The transformation to a low-dimensional latent space forces the autoencoders to retain the dominant features alone. The convolutional autoencoder used in this approach does exactly that by de-noising the  noisy microstructure input. The output of the convolutional encoder is almost in its pure form, free from noise and therefore enables the DeepONet to make stable predictions. The decoder accurately reconstructs the microstructure from the predictions made by DeepONet in the latent space.

\begin{table}[H]
    \begin{center}
        \caption{Prediction errors with noisy inputs}
        \label{table: noise}
        \begin{tabular}{cc}
            \hline
            \hline
            \% Noise level & $\mathcal{D}_{\rm test}$ \\
            \hline
            \hline \\
            0.5            & 0.04344 \\
            1              & 0.04314 \\
            2              & 0.04260 \\
            3              & 0.04282 \\
            4              & 0.04300 \\
            5              & 0.04364 \\
            10             & 0.04738 \\
            \hline
        \end{tabular}
    \end{center}
    
\end{table}

\clearpage

\section*{Supplementary Note 6: Integrating DeepONet with MEMPHIS solver}

We devise a hybrid model that utilizes autoencoder-DeepONet architecture to leap in time combined with the high-fidelity simulations.
The strategy implemented and the results on test data have been provided in Supplementary \autoref{fig: integration_results_test}. The detailed algorithm has been provided in the main manuscript.

\begin{figure}[H]
    \centerline{\includegraphics[height=0.80\textheight]{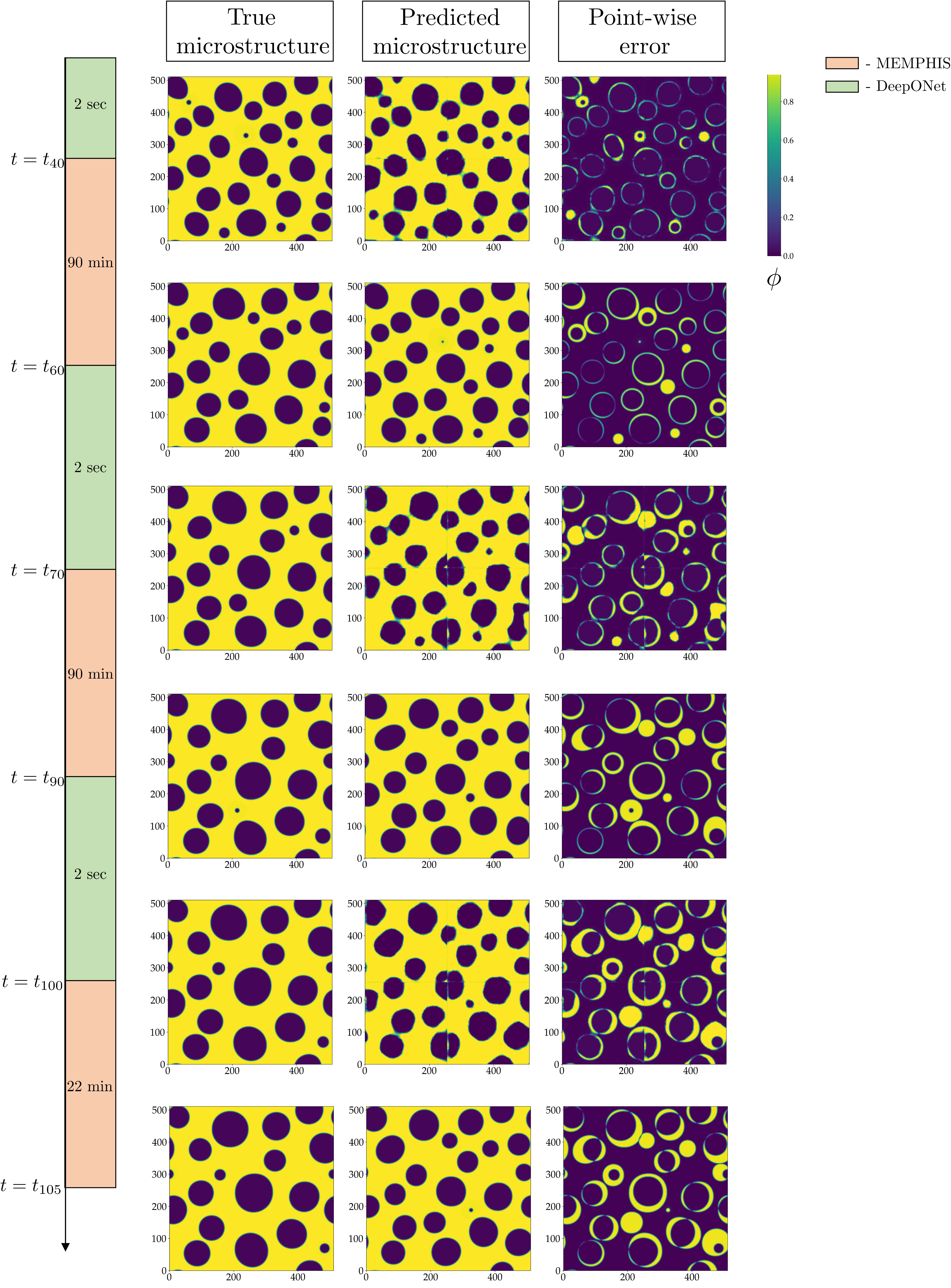}}
    \caption{{\bf Schematic of our hybrid approach for integrating DeepONet with the numerical solver MEMPHIS to accelerate phase-field predictions.}
    The computational time corresponding to the autoencoder--DeepONet model and the MEMPHIS solver for one realization in the test dataset is reported in this figure. The error is shown on the third column.}
    \label{fig: integration_results_test}
\end{figure}

\end{document}